\documentclass[article]{JHEP3}
\usepackage{amsmath,amssymb}
\usepackage{cite}
\usepackage{float}
\usepackage{mathtools}
\usepackage{subcaption}
\usepackage{comment}
\usepackage{amsfonts}
\usepackage{mathrsfs}
\usepackage[utf8]{inputenc}
\usepackage{graphicx}
\newtheorem{Conjecture}{Conjecture}
\DeclareMathOperator\arctanh{arctanh}

\allowdisplaybreaks

\title{Fields on accelerating and super accelerating black holes: stability, quasinormal modes, greybody factors and strong cosmic censorship}
\author{Francisco Silva$^a$
\\
$^{a}$
Departamento de F\'\i sica,\\ Faculdade de Ci\^encias da Universidade de Lisboa,\\
Campo Grande, 1749-016 Lisboa, Portugal\\
\email{fc58297@alunos.ciencias.ulisboa.pt}
}

\abstract{We investigate the propagation of massless scalar, Dirac, and Maxwell fields on accelerating and super accelerating black hole spacetimes described by the C-metric. The latter regime has only recently been identified and remains largely unexplored. We study the linear and mode stability of these backgrounds and derive analytical expressions for the quasinormal mode spectra and greybody factors in various spacetime limits. We classify the quasinormal mode families, paying particular attention to new families arising in the super accelerating regime. We also investigate strong cosmic censorship through Christodoulou's $\beta$ criterion, finding regions of parameter space where it may be violated, including in the super accelerating regime. These violations are possible for massless scalar and Maxwell fields, whereas Dirac fields preserve the conjecture.}

\begin{document}

\maketitle

\vfill

\eject

%%%%%%%%%%%%%%%%%%%%%%%%%%%%%%%%%%%%%%%%%%%%%%%%%%%%%%%%%%%%%%%%%%%%%%
%%%%%%%%%%%%%%%%%%%%%%%%%%%%%%%%%%%%%%%%%%%%%%%%%%%%%%%%%%%%%%%%%%%%%%
\section{Introduction}\label{Introduction}
%%%%%%%%%%%%%%%%%%%%%%%%%%%%%%%%%%%%%%%%%%%%%%%%%%%%%%%%%%%%%%%%%%%%%%
%%%%%%%%%%%%%%%%%%%%%%%%%%%%%%%%%%%%%%%%%%%%%%%%%%%%%%%%%%%%%%%%%%%%%%
\noindent
The study of accelerating black holes has attracted sustained interest in the General relativity (GR) community. These are found to be described by the C-metric \cite{Kinnersley:1970zw}, which provides a way to include acceleration in a black hole spacetime. Physically, recent observations motivate the consideration of astrophysical environments where black holes reside or are formed in dense environments like globular clusters or accompanied by other massive objects. Additionally, there is a phenomenological effect in which the emission of gravitational waves by black hole binary systems, during the inspiral phase, tend to have a preferred direction, resulting in a recoil acceleration of the remnant black hole. This process is known as the black hole kick \cite{Bruegmann:2007bri}. One hopes, therefore, that these accelerating black holes can be, if not identically, approximately described by the C-metric.

Spectroscopic quantities such as quasinormal modes (QNMs), the greybody factor (GBF) and the shadow radius of accelerating black holes have been subject to recent works \cite{Frost:2020zcy, Moura:2026ken}. These provide different possible observational channels that can probe the black hole parameters and compare strong field gravity predictions with observations. Not only that, but quasinormal modes also provide us with important information regarding the black hole's stability.

Recently, a new regime of the C-metric has been identified, called the super accelerating regime \cite{Silva:2026tev}. This regime is found when the acceleration parameter is such that the horizon labeled by acceleration ($r_A$), becomes either the event horizon or Cauchy horizon of the black hole, opposed to the regular accelerating regime, where $r_A$ describes only the cosmological horizon. Moreover, new properties on the spacetime such as a new axis of symmetry, its Komar mass and new near extremal configurations have been discussed in \cite{Silva:2026tev}.

Fields on the super accelerating regime, however, have not yet been previously studied; therefore it is the goal of this work to do so. While spectroscopy quantities, such as QNM and the GBF can indeed be computed for this regime as well, to use them to probe astrophysical black holes is not the main motivation of this work, as very highly accelerating black holes are hard to conceive in astrophysical scenarios and possibly deemed to be unphysical. Needless to say, our aim here is to study the super accelerating regime as a mathematical solution of Einstein's equations. Indeed, more specifically, we are interested in knowing, whether it is stable as a solution and whether we can find new results regarding the strong cosmic censorship (SCC) conjecture.

To briefly introduce the SCC conjecture to the unfamiliar reader, let us take a step back and consider the role of Cauchy horizons in GR. These horizons are generically thought to be unstable whose source of instability is due to the infinite blueshift characteristic of them. Moreover, were they not unstable, a lack of predictability within GR would emerge, as these horizons form the boundary of the maximal Cauchy development. Therefore, we are usually glad that this instability exists, as this preserves predictability within GR. The modern version of the strong cosmic censorship statement is due to Christodoulou \cite{Christodoulou:2008nj}
\begin{Conjecture}[Christodoulou Version of Strong Cosmic Censorship]
    Generically, it is not possible to extend the maximal Cauchy development across the Cauchy Horizon as a weak metric solution of the equations of motion.
    \label{SCCconjecture}
\end{Conjecture}
The key point here is the notion of a \emph{weak solution}. Earlier formulations of strong cosmic censorship, such as Penrose's original $C^0$ formulation, or other formulation based on $C^2$-inextendibility, impose a different notion of regularity. The distinction is important, because ultimately, we wish to impose the required level of regularity on the Cauchy horizon, such that an observer cannot cross it and thus preserving causality. Christodoulou's version of strong cosmic censorship, thus, seems to be optimal as it includes, for instances, physically motivated solutions such as shock waves \cite{Dias:2018etb}, but is not as restrictive as the $C^2$ version and not as allowing as Penrose's $C^0$ version which has already been proven false \cite{Dafermos:2017dbw}.

As we will see in section \ref{Stability of the Cauchy horizon}, Christodoulou later came up with a criterion that allows us to evaluate whether the given conjecture is true. These assessments have been performed in the regular accelerating regime for scalar fields in \cite{Destounis:2020yav}. In this paper, we will see how this generalizes to the super accelerating regimes, where the horizon labeled by $r_A$ plays a different role from the cosmological horizon, and we will study how different spin perturbations enter in the evaluation of the conjecture as well.

The structure of this work is the following. In section \ref{C-metric}, we review the regular and super accelerating regimes of the C-metric. Later, in section \ref{Fields}, we study massless scalar, Dirac and Maxwell fields living on the background spacetime of the black hole. Afterwards, in section \ref{Linear stability}, we analyze the linear stability of these fields, with particular interest on the super accelerating regimes. From then on, we consider the quasinormal mode problem, but first review a known proof showing that there are no mode instabilities on the black hole, in section \ref{Mode stability}. In section \ref{Quasinormal mode families}, we classify all the black hole mode families and provide analytical solutions for the QNM frequency spectra and the greybody factor. %Given that, when quasinormal modes can be analytically computable, so should greybody factors, therefore these quantities are also computed. %This completes the work of \cite{Destounis:2020pjk}, as it goes through the actual analytical proof these expressions as well as discusses the super accelerating regime scenarios. 
Finally, in section \ref{Stability of the Cauchy horizon}, we use the analytical expressions obtained for the QNM frequencies to analyze whether strong cosmic censorship is preserved on both accelerating and super accelerating black holes, by use of Christodoulou's $\beta$ criterion on strong cosmic censorship.

\section{Accelerating and super accelerating black holes: The C-metric}\label{C-metric}
%%%%%%%%%%%%%%%%%%%%%%%%%%%%%%%%%%%%%%%%%%%%%%%%%%%%%%%%%%%%%%%%%%%%%%
%%%%%%%%%%%%%%%%%%%%%%%%%%%%%%%%%%%%%%%%%%%%%%%%%%%%%%%%%%%%%%%%%%%%%%
\noindent
Since the work of Kinnersley and Walker \cite{Kinnersley:1970zw}, the C-metric has been interpreted as describing uniformly accelerating black holes. In the spherical-like coordinates introduced by Hong and Teo \cite{Hong:2003gx}, it is expressed as 
\begin{equation}
    ds^2=\frac{1}{\Omega^2(r,\theta)}\left(-f(r)dt^2+\frac{dr^2}{f(r)}+\frac{r^2d\theta^2}{P(\theta)}+r^2P(\theta)\sin^2{\theta}d\varphi^2\right),
    \label{Metric}
\end{equation}
where metric functions are defined by
\begin{equation}
    f(r)=\left(1-\frac{2M}{r}+\frac{Q^2}{r^2}\right)\left(1- A^2r^2\right)-\frac{\Lambda}{3}r^2, \quad P(\theta)= 1-2AM\cos{\theta}+A^2Q^2\cos^2{\theta},
\label{metric functions}
\end{equation}
\begin{equation}
    \Omega(r,\theta)=1-Ar \cos{\theta}.
    \label{conformal factor}
\end{equation}
The parameters $M,Q$ and $A$ represent the black hole's mass, charge and acceleration parameters. They should not be simply interpreted as the black holes real mass and charge, as these quantities demand a careful definition \cite{Griffiths:2006tk, Silva:2026tev}. 

Moreover, the electromagnetic vector potential $\mathscr{A}$ and its associated field strength $F$ that realize this black hole solution are given by
\begin{equation}
    \mathscr{A}=-\frac{Q}{r}dt, \quad F=\frac{Q}{r^2}dr\wedge dt.
\end{equation}

The horizons of this black hole are dictated by the zeros of the function $f(r)$ and therefore are given by
\begin{equation}
    r_\pm=M\pm \sqrt{M^2-Q^2},\quad r_{\pm A}=\pm\frac{1}{A}.
\end{equation}
As a consequence, the horizons of the Reissner Nordstr\"{o}m black hole are preserved for a black hole with acceleration $A$. This naturally imposes that $M>|Q|$. On the other hand, it seems that there would be no limit on the acceleration parameter $A$. Consequently, it would be possible to have different accelerating regimes. These are defined by, first
\begin{equation}
    r_-<r_+<r_A,
\end{equation}
which we call the regular accelerating regime, and
\begin{equation}
    r_-<r_A<r_+\quad \text{or}\quad r_A<r_-<r_+.
\end{equation}
which we call the first and second super accelerating regimes respectively. There are two points to emphasize here. One is that the region in between the outer horizon and the cosmological horizon is timelike in $r$ for any of the accelerating regimes defined above\footnote{We opt to classify the horizons in an ascending order, as the inner horizon, the outer horizon and the cosmological horizon. Alternatively we will also call it the Cauchy horizon and event horizon respectively for the first two.}. The other is that for the super accelerating regimes, there is now a region in which $P(\theta)\leq 0$. In order to preserve the metric signature, we restrict the usual $(0,\pi)$ range in which the coordinate $\theta$ takes values to a smaller subset. For the first super accelerating regime, the region where $P(\theta)<0$ is defined by $\theta \in (0,\theta_+)$, where $\theta_+=\arccos(r_A/r_+)$. For the second super accelerating regime it is in between $\theta \in (\theta_-,\theta_+)$, where $\theta_-=\arccos(r_A/r_-)$. Consequently the observer should live on the $\theta \in (\theta_+,\pi)$ region for the first super accelerating regime and on either $\theta \in (0,\theta_-)$ or $\theta\in(0,\theta_+)$ for the second super accelerating regime. At $\theta=\theta_\pm$ the function $P(\theta)=0$. This signals that the azimuthal Killing vector $\partial_\varphi$ has zero norm there and therefore this boundary can be interpreted as a new axis of symmetry. Not only that, but one can also prove that the resulting angular spacetime manifold is conformal to the topologically equivalent of a two-sphere \cite{Silva:2026tev}. Additionally, it is important to highlight that there are in general two conical singularities in the causal region of the spacetime, representing points where an infinitesimal small circle has circumference different from $2\pi \times \text{Radius}$. These have the interpretation of being cosmic strings/struts that provide acceleration to the black hole. On the regular accelerating regime, they live on the $\theta=0$ and $\theta=\pi$ regions, while for the super accelerating black holes they are present on $\theta=\theta_\pm$ as well. A more detailed discussion regarding the horizons structure and conical singularities on the super accelerating regime can be found in \cite{Silva:2026tev}.

\section{Fields on the background metric}\label{Fields}
%%%%%%%%%%%%%%%%%%%%%%%%%%%%%%%%%%%%%%%%%%%%%%%%%%%%%%%%%%%%%%%%%%%%%%
%%%%%%%%%%%%%%%%%%%%%%%%%%%%%%%%%%%%%%%%%%%%%%%%%%%%%%%%%%%%%%%%%%%%%%
\noindent
In this section, we will discuss how fields behave on the C-metric curved background. In general there are two approaches to this problem. The first is based on the Newman–Penrose formalism, in which one studies perturbations of gauge-invariant quantities, by the construction of a Teukolsky-like equation. The second consists of analyzing the equations of motion of the corresponding fields. For instance, one considers the Klein–Gordon equation for spin-$0$ fields, the Dirac equation for spin-$1/2$ fields, and Maxwell's equations for spin-$1$ fields.

A Teukolsky-like equation of motion for fields on the C-metric has been obtained by Kofro\v{n} in \cite{Kofron:2015gli} and for rotating accelerating black holes in \cite{Kofron:2016dyk}. However, in this paper will be concerned with the second approach. 

Before starting calculating mechanically, notice that the metric \eqref{Metric} has a conformal factor which depends on the $r$ and $\theta$ coordinates. In general, it is expected that this complicates the separability of the equations of motion. Therefore, it is useful to consider conformally invariant fields. This excludes massive fields and the graviton as well. As such, we are left with the massless scalar, Dirac, Maxwell and Rarita-Schwinger fields. However, we will not consider the latter field in this work, although still interesting to explore. Additionally, we will take the test field approximation, where there is no backreaction on the background spacetime. It is relevant to mention that although the massless scalar field has been studied in many works \cite{Destounis:2022rpk, Destounis:2020pjk, Destounis:2020yav, BarraganAmado:2023wxt, Moura:2026ken, Lei:2023mqx}, the spin $1/2$ and spin $1$ field have been subject of less study on the C-metric spacetime. Their field dynamics has been considered in \cite{Kofron:2015gli, Kofron:2016dyk}, but on the study of QNMs, GBFs and the stability of the Cauchy horizon, these fields have been mostly neglected.

If the readers are either familiar or uninterested on how the equations of motion for these fields are obtained in the black hole's curved background, they may skip to subsection \ref{On the boundary conditions} to progress with the paper's results.

\subsection{The scalar field}
%%%%%%%%%%%%%%%%%%%%%%%%%%%%%%%%%%%%%%%%%%%%%%%%%%%%%%%%%%%%%%%%%%%%%%
%%%%%%%%%%%%%%%%%%%%%%%%%%%%%%%%%%%%%%%%%%%%%%%%%%%%%%%%%%%%%%%%%%%%%%
\noindent
The equations of motion governing the massless charged scalar field were found to be separable on the C-metric \cite{Destounis:2022rpk}. Here we will rederive the corresponding equations of motion for an uncharged field. 

Consider $\Phi$ a massless scalar field. Its corresponding equations of motion are obtained by applying the curved spacetime D'Alembertian operator $\square_g$ on $\Phi$ and setting it to zero
\begin{equation}
    \square_g \Phi =0.
    \label{KG}
\end{equation}

However, as argued before, the trick to separating this equation is by performing a conformal transformation; both on the field and on the metric at the same time. It is known that under the transformations 
\begin{equation}
    \Phi \rightarrow \tilde{\Phi}= \Omega^{-1}\Phi,\quad g_{\mu \nu} \rightarrow \tilde g_{\mu \nu}=\Omega ^{2}g_{\mu \nu}
\end{equation}
equation \eqref{KG} transforms to 
\begin{equation}
    \Box_{\tilde g}\tilde \Phi -\frac{1}{6}\tilde{R}\tilde{\Phi}=0,
    \label{conformal KG}
\end{equation}
where $\tilde{R}$ is the Ricci scalar of the conformally rescaled metric. It turns out that this object is just the sum of an $r$ dependent function plus a $\theta$ dependent function. This is the key identity that allows for the separability of the equations,

\begin{equation}
    \tilde R =- (r^2f''(r)+4rf'(r)+2f(r)+P''(\theta)+3\cot{\theta}P'(\theta)-2P(\theta)).
    \label{Ricci}
\end{equation}
We introduce the following ansatz
\begin{equation}
    \tilde{\Phi}=e^{-i\omega t}e^{im\varphi}\frac{\phi_0(r)}{r}\chi_0(\theta),
    \label{scalar ansatz}
\end{equation}
where $\omega$ is the field frequency, $m$ is the azimuthal angular number and $\phi_0$, $\chi_0$ are the corresponding radial and angular functions governing the field dynamics.
Plugging \eqref{scalar ansatz} onto \eqref{conformal KG}, we find that the resulting radial equation of motion is Schr\"{o}dinger-like
\begin{equation}
    \left(-\frac{d^2}{dx^2}+\,_0V_{r}\right) \phi_0=\omega^2\phi_0
    \label{finalradialscalareq}
\end{equation}
with
\begin{equation}
    \,_0V_r=f(r)\left(\frac{\lambda_0-1/3+A^2Q^2/3}{r^2}+\frac{2M}{r^3}-\frac{2Q^2}{r^4} \right),
    \label{finalradialscalarpotential}%expandVr
\end{equation}
and the radial tortoise coordinate ($x$) is defined by
\begin{equation}
dx=\frac{dr}{f(r)}.
\label{tortoise}
\end{equation}
The $\lambda_0$ in \eqref{finalradialscalarpotential} is a separation constant introduced when the angular and radial equations are separated. The corresponding motion of the angular sector is thus governed by
\begin{equation}
    \frac{d^2\chi_0}{dz^2}-(m^2-\,_0V_\theta)\chi_0(\theta)=0,
    \label{finalangularscalareq}
\end{equation}
where 
\begin{equation}
    _0V_\theta=P(\theta)\left((\lambda_0-1/3+A^2Q^2/3)\sin^2\theta +2AM\cos{\theta}\sin^2{\theta} -2A^2Q^2\cos^2\theta\sin^2\theta\right).
    \label{finalangularscalarpotential}%expandvt
\end{equation}
and the angular tortoise coordinate ($z$) is defined as
\begin{equation}
    dz=\frac{d\theta}{P(\theta) \sin\theta}.
    \label{angular tortoise}
\end{equation}

The respective values that $\lambda_0$, $\omega$ and $m$ can take will be discussed more deeply and generically for the different spin fields when we reach subsection \ref{On the boundary conditions}. For now let us proceeded to analyze the Dirac field.

\subsection{The Dirac field}\label{The Dirac field}
%%%%%%%%%%%%%%%%%%%%%%%%%%%%%%%%%%%%%%%%%%%%%%%%%%%%%%%%%%%%%%%%%%%%%%%%%%%%%%%%%%%
%%%%%%%%%%%%%%%%%%%%%%%%%%%%%%%%%%%%%%%%%%%%%%%%%%%%%%%%%%%%%%%%%%%%%%
\noindent
The Dirac equation is responsible on the dynamics of spin $1/2$ fermions. For a massless fermion it is expressed by
\begin{equation}
    i\gamma^\mu \nabla_\mu \psi =0,
    \label{Dirac equation}
\end{equation}
where $\psi$ is a four component spinor, $\nabla_\mu$ is the spinor covariant derivative and $\gamma^\mu$ are the curved spacetime gamma matrices.

Once again, we shall perform a conformal transformation, however it is convenient to introduce an $r$ factor together with the conformal factor of the metric. This leads to
\begin{equation}
    \psi \rightarrow \tilde{\psi}= r^{3/2}\Omega^{-3/2}\psi,\quad g_{\mu \nu} \rightarrow \tilde g_{\mu \nu}=r^{-2}\Omega ^{2}g_{\mu \nu},\quad \gamma^\mu \rightarrow \tilde \gamma^\mu, \quad \gamma^\mu \nabla_\mu \psi \rightarrow r^{-5/2}\Omega^{5/2}\tilde \gamma^\mu \tilde \nabla_\mu \tilde \psi,
    \label{Dirac transformation}
\end{equation}
where we note by the last identity in equation \eqref{Dirac transformation} that the massless Dirac equation \eqref{Dirac equation} is invariant under conformal transformations. By \eqref{Dirac transformation}, the metric factorizes into the metric of a two-dimensional Lorentizan manifold with coordinates $(t,r)$ and a 2d compact manifold with coordinates $(\theta,\varphi)$. We remind ourself that a massless Dirac field is equivalent to two decoupled Weyl fermions and therefore introduce the ansatz 

\begin{equation}
\tilde \psi=\phi^{(+)}(r,t)\otimes\chi^{(+)}(\theta,\varphi)+\phi^{(-)}(r,t)\otimes\chi^{(-)}(\theta,\varphi),
\label{fermionansatz}
\end{equation}
where the $\pm$ fields can be thought as $2$ spinors or equivalently as $4$ spinors whose only non vanishing components are the first two for $(+)$ and and the last two for $(-)$.

With all of this, we can expand the Dirac equation \eqref{Dirac equation} to 
\begin{equation}
    \left[ (\tilde\gamma^t\tilde\nabla_t+\tilde\gamma^r\tilde\nabla_r) \otimes \mathbb{I}+\tilde \gamma^5 \otimes  (\tilde \gamma^j\tilde\nabla_j) \right]\tilde \psi=0,
    \label{Diraceq2}
\end{equation}
where $j=(\theta, \phi)$ and $\tilde \gamma^5=i\tilde \gamma ^0$. 

We are now ready to introduce the ansatz 
\begin{equation}
    \tilde \psi =\phi_{1/2}^{(\pm)}(t,r)\Upsilon ^{\pm}(\theta,\varphi)
    \label{fermiontrthetaphi}
\end{equation}
Using equation \eqref{fermionansatz} on \eqref{Diraceq2}, together with the ansatz in \eqref{fermiontrthetaphi}, we find out that the angular and radial part of the equations separate.
The separation constant $\lambda_{1/2}$ is introduced, such that the angular equation becomes
\begin{equation}
\tilde \gamma ^j\nabla_j \Upsilon^{(\pm)}=\pm i \lambda_{1/2}\Upsilon^{(\pm)}.
\label{fermionangular}
\end{equation}

We will now focus our attention on the angular sector. Notice that working with $\Upsilon^{(+)}$ is equivalent to working with $\Upsilon^{(-)}$, so let us consider $\Upsilon^{(+)}$ without loss of generality. We will now expand \eqref{fermionangular}, so that the resulting equations are more easily dealt with. The covariant derivative is given by
\begin{equation}
    \nabla _\mu \Upsilon ^{(+)}=\partial_\mu \Upsilon^{(+)} +\frac{1}{4}\omega^{ab}_\mu \tilde \gamma_{ab}\Upsilon^{(+)},
\end{equation}
where the spin connection $\omega^{ab}_\mu$ is defined through a vielbein $e^a_\mu$, such that 
\begin{equation}
    g_{\mu \nu}=e^a_\mu e^b_\nu \eta _{ab}\quad \text{and}\quad \omega^{ab}_\mu = e^{a\nu}(\partial_\mu e^b_\nu +\Gamma ^\nu _{\mu \rho}e^b_\rho).
\end{equation}
The Latin letters now refer to a local orthonormal frame and the Greek letters refer to spacetime indices. $\eta _{ab}$ is the Minkowskian metric (on this case however, a two dimensional euclidean metric), $\Gamma ^\nu _{\mu \rho}$ are the Christoffel symbols and $\tilde \gamma_{ab}=\frac{1}{2}[\tilde \gamma_a,\tilde \gamma_b]$ is the commutator of the gamma matrices.

A natural choice for the vielbein is
\begin{equation}
    e^1= \frac{d\theta}{\sqrt{P(\theta)}}\quad \text{and}\quad  e^2=\sqrt{P(\theta)}\sin \theta d\varphi.
\end{equation}
Therefore, the curved spacetime gamma matrices are given by
\begin{equation}
    \tilde \gamma^\mu =e^\mu _a\tilde  \gamma^a,
\end{equation}
while the flat gamma matrices (since we are considering $2$ spinors only) can be taken to be
\begin{equation}
    \tilde \gamma^1= \sigma^1= \begin{pmatrix}
        0 & 1 \\
        1 & 0
    \end{pmatrix},
    \quad \tilde \gamma^2=\sigma^2=\begin{pmatrix}
        0 & -i \\
        i & 0
    \end{pmatrix},
\end{equation}
where $\sigma^i, i=1,2,3$ denotes the Pauli matrices. The third Pauli matrix is defined as 
\begin{equation}
    \sigma^3=\begin{pmatrix}
        1 & 0 \\
        0 & -1
    \end{pmatrix}.
\end{equation}
We now need the spin coefficients. We find that the only non vanishing ones are 
\begin{equation}
    \omega ^{12}_\varphi = -\omega ^{21}_\varphi=-\cos \theta P(\theta) -\frac{1}{2}\sin{\theta} P'(\theta)
\end{equation}
and since $\tilde \gamma_{12}=\frac{1}{2}[\tilde \gamma_1,\tilde \gamma_2]=i\sigma^3$, we can finally conclude that the angular Dirac operator can be written as
\begin{equation}
    \tilde \gamma^i \nabla_i=\tilde \gamma^\theta\nabla_\theta+\tilde \gamma^\varphi\nabla_\varphi=\sqrt{P}\sigma^1\frac{\partial}{\partial\theta}+\frac{\sigma^2}{\sqrt{P}\sin\theta}\left(\frac{\partial}{\partial\varphi} -\left(\frac{P}{2}\cos{\theta}+\frac{1}{4}\frac{P'}{\sqrt{P}}\right)i\sigma^3 \right),
\end{equation}
where the prime corresponds to derivation with respect to $\theta$. We further separate the azimuthal and polar coordinates by introducing $\Upsilon^{(+)}(\theta,\varphi)=e^{im\varphi}\chi_{1/2}^{(+)}(\theta)$, where we obtain 

\begin{equation}
    \left(\sqrt{P(\theta)}\frac{d}{d\theta}+\frac{\sqrt{P(\theta)}}{2}\cot{\theta}+\frac{P'(\theta)}{4\sqrt{P(\theta)}}\right)\sigma^1 \chi_{1/2}^{(+)}(\theta) +\frac{im}{\sqrt{P(\theta)}\sin{\theta}} \sigma^2\chi_{1/2}^{(+)}(\theta)= i\lambda_{1/2}\mathbb{I}\chi_{1/2} ^{(+)}(\theta).
    \label{angulardirac}
\end{equation}
Notice the identity
\begin{equation}
    \left[\frac{d}{d\theta}+\frac{1}{2}\cot{\theta}+\frac{1}{4}\frac{P'(\theta)}{P(\theta)}\right]\chi_{1/2}^{(+)}=\left(\sin{\theta}\sqrt{P(\theta)}\right)^{-1/2}\frac{d}{d\theta}\left(\left(\sin{\theta}\sqrt{P(\theta)}\right)^{1/2}\chi_{1/2}^{(+)}\right).
\end{equation}
This motivates us to introduce the ansatz
\begin{equation}
    \chi_{1/2}^{(+)}=\left(\sin \theta \sqrt{P(\theta)}\right)^{-1/2}\begin{pmatrix}
        i\mathscr{H} \\
        \mathscr{J}
    \end{pmatrix}.
\end{equation}
We further use the angular tortoise coordinate $z$ defined in equation \eqref{angular tortoise} and plug this into equation \eqref{angulardirac} giving the following two first order coupled differential equations

\begin{equation}
    \begin{cases}
            \left(\frac{d}{dz}-m\right)\mathscr{H}=\lambda_{1/2}\sqrt{P}\sin{\theta}\mathscr{J}\\
    \left( \frac{d}{dz}+m\right)\mathscr{J}=-\lambda_{1/2}\sqrt{P}\sin\theta\mathscr{H}.
    \end{cases}
    \label{fermion pair dif eq}
\end{equation}
However, we wish to find a single second order differential equation. In order to do that, we introduce 
\begin{equation}
    \mathscr{K}=\sqrt{P}\sin\theta.
\end{equation}
It now becomes clear that the system \eqref{fermion pair dif eq} yields
\begin{equation}
    \mathscr{H}=-\frac{1}{\lambda_{1/2}\mathscr{K}}\left(\frac{d}{dz} +m\right)\mathscr{J} \quad \text{and}\quad \mathscr{J}=\frac{1}{\lambda_{1/2}\mathscr{K}}\left(\frac{d}{dz}-m \right)\mathscr{H}.
    \label{fermion3}
\end{equation}
As such, plugging this again on the system of equations \eqref{fermion pair dif eq}, we find a single order differential equation for both $\mathscr{H}$ and $\mathscr{J}$. It suffices to consider $\mathscr{H}$.
\begin{equation}
    \frac{d^2\mathscr{H}}{dz^2}-\frac{\frac{d}{dz}\mathscr{K}}{\mathscr{K}}\frac{d\mathscr{H}}{dz}+\left(\lambda_{1/2}^2\mathscr{K}^2 -m^2+m\frac{\frac{d}{dz}\mathscr{K}}{\mathscr{K}}\right)\mathscr{H}=0.
    \label{fermion4}
\end{equation}
It is now possible to do a consistency check. By turning off acceleration, we ultimately get $\mathscr{K}=\sin{\theta}$ and \eqref{fermion4} will give the spin weighted spherical harmonics. Another possible way to validate \eqref{fermion4} is to check if the Lichnerowicz formula holds
\begin{equation}
    (\tilde \gamma^i\tilde \nabla_i)^2=\tilde \nabla^2+\frac{\tilde R}{4},
\end{equation}
where $\tilde \nabla^2$ is the spinor Laplacian (Laplacian with the spin connection) and $\tilde R$ the Ricci curvature scalar on the conformal compact manifold defined by the $\theta$ and $\varphi$\footnote{The Ricci scalar mentioned here is given in \eqref{Ricci} but taking $f(r)=0$.}.

To conclude the treatment of the angular part, we wish to write \eqref{fermion4} in the standard Schrödinger-like form. To do so we introduce the field 
\begin{equation}
    \mathscr{E}=\sqrt{\mathscr{K}}\mathscr{H},
\end{equation}
and now equation \eqref{fermion4} becomes
\begin{equation}
    \frac{d^2\mathscr{E}}{dz^2}-\left(m^2-
    \,_{1/2}V_\theta \right)\mathscr{E}=0,
    \label{finalangulardiraceq}
\end{equation}
with
\begin{equation}
_{1/2}V_\theta=\lambda_{1/2}^2\mathscr{K}^2+m\frac{\frac{d}{dz}\mathscr{K}}{\mathscr{K}}+\frac{\frac{d^2}{dz^2}\mathscr{\mathscr{K}}}{2\mathscr{K}}-\frac{3}{4}\left(\frac{\frac{d}{dz}\mathscr{K}}{\mathscr{K}}\right)^2.
\label{finalangulardiracpotential}
\end{equation}
This, therefore, finishes the treatment of the angular equation.

We shall now look into the radial part. We fall back to equation \eqref{fermiontrthetaphi} where we separated the angular and radial plus time parts of the $\tilde \psi$ field. The treatment of $\phi^{(+)}$ will be equivalent to the treatment of $\phi^{(-)}$, so let us only consider the first one without loss of generality. Indeed, the procedure now is quite similar as the one we have developed for the angular equation so briefly go into the details. By plugging \eqref{fermiontrthetaphi} into \eqref{Diraceq2} the radial and time equation now obey 
\begin{equation}
    \sigma^2\left(\sqrt{f(r)}r\right)\left[ \frac{\partial}{\partial r}+\frac{r}{2\sqrt{f(r)}}\frac{d}{dr}\left(\frac{\sqrt{f(r)}}{r} \right)\right]\phi^{(+)}-i\sigma^1 \lambda_{1/2}\phi^{(+)}=i\sigma^{3}\left( \frac{r}{\sqrt{f(r)}}\right)\frac{\partial \phi^{(+)}}{\partial t}.
    \label{diracrt}
\end{equation}
We now introduce the clever ansatz
\begin{equation}
    \phi^{(+)}=\left( \frac{\sqrt{f(r)}}{r}\right)^{-1/2} e^{-i\omega t} \begin{pmatrix}
        i \mathscr{G}(r) \\
        \mathscr{F}(r)
    \end{pmatrix},
\end{equation}
and plug it into equation \eqref{diracrt}. In fact, after some algebra, the set of first order coupled differential equations of motion become
\begin{equation}
\begin{cases}
    f(r)\frac{d\mathscr{G}(r)}{dr}-\frac{\sqrt{f(r)}}{r}\lambda_{1/2}\mathscr{G}(r)=\omega \mathscr{F}(r), \\
    f(r)\frac{d\mathscr{F}(r)}{dr}+\frac{\sqrt{f(r)}}{r}\lambda_{1/2}\mathscr{F}(r)=-\omega \mathscr{G}(r).
\end{cases}
\end{equation}
This can be simplified by introducing the superpotential 
\begin{equation}
    \mathscr{W}=\lambda_{1/2}\sqrt{f(r)}/r
\end{equation}
and the tortoise coordinate defined in \eqref{tortoise}. It yields 
\begin{equation}
    \begin{cases}
    \left(\frac{d}{dx}-\mathscr{W}\right)\mathscr{G}=\omega \mathscr{F}, \\
    \left(\frac{d}{dx}+\mathscr{W}\right)\mathscr{F}=-\omega \mathscr{G}.
    \end{cases}
\end{equation}
Differentiating both sides by the tortoise coordinate $x$, gives the following simplification
\begin{equation}
    \left(-\frac{d^2}{dx^2}+\,_{1/2}V_{r}^+\right)\mathscr{F}=\omega^2\mathscr{F}\quad \text{and}\quad 
    \left(-\frac{d^2}{dx^2}+\,_{1/2}V_{r}^-\right)\mathscr{G}=\omega^2\mathscr{G},
    \label{finalradialdiraceq}
\end{equation}
where \begin{equation}
    \,_{1/2}V_{r}^\pm =\mathscr{W}^2\pm \frac{d\mathscr{W}}{dx}=\frac{\lambda_{1/2}^2f(r)}{r^2}\pm \lambda_{1/2}\sqrt{f(r)}\left( \frac{f'(r)}{2r}-\frac{f(r)}{r^2}\right).
    \label{finalradialdirapotential}
\end{equation}
These are the equations governing the radial fermionic modes. It is curious and interesting to see that they have an underlying supersymmetric structure. Consequently, it is known that for nonzero $\omega$, the partner $\mathscr{F}$ and $\mathscr{G}$ modes will have the same spectrum. They are thus called isospectral.

\subsection{The Maxwell field}
%%%%%%%%%%%%%%%%%%%%%%%%%%%%%%%%%%%%%%%%%%%%%%%%%%%%%%%%%%%%%%%%%%%%%%
%%%%%%%%%%%%%%%%%%%%%%%%%%%%%%%%%%%%%%%%%%%%%%%%%%%%%%%%%%%%%%%%%%%%%%
\noindent
Let us now consider electromagnetic (Maxwell) fields.
The Maxwell field equations are known to be conformally invariant in the four spacetime dimensions. In vacuum, these are given by
\begin{equation}
    \nabla_\mu F^{\mu \nu}=0,\quad \text{where} \quad F_{\mu \nu}=\partial_\mu \mathscr{A}_\nu -\partial_\nu \mathscr{A}_\mu.
    \label{MaxwellEQ}
\end{equation}
Our black hole is charged and therefore, there is a background electric field 
\begin{equation}
    \mathscr{A}_\mu ^{(0)}=\left(-\frac{Q}{r},0,0,0\right),
\end{equation}
hence we consider perturbations $\delta A_\mu$ on this background such that 
\begin{equation}
    \mathscr{A}_\mu = \mathscr{A}_\mu^{(0)}+\delta \mathscr{A}_\mu.
\end{equation}

To parametrize the perturbations, we are inspired by the Hodge decomposition to write the general form of the perturbation as 
\begin{equation}
\delta \mathscr{A}_\mu = \delta \mathscr{A}_\mu ^{P}+\delta \mathscr{A}_\mu ^{{A}},
\end{equation}
where $P$ stands for polar and $A$ stands for axial.
Focusing on the angular manifold whose coordinates are labeled by $a=(\theta, \varphi)$, Hodge theorem tells us that these perturbations are given by 
\begin{equation}
    \delta \mathscr{A}^P_a=\nabla_ {a}\Pi,\quad \delta \mathscr{A}^A_a={\epsilon_a}^b \nabla_b \Psi,
\end{equation}
where $\Pi$ and $\Psi $ are some generic fields and $\nabla_a$ is the covariant derivative on the respective angular manifold. The reason why a harmonic term is not included in the Hodge decomposition is related with the fact that in a manifold topologically equivalent to the two-sphere there are no non trivial harmonic forms; the manifold really is topologically equivalent to a two-sphere, so this result holds. We separate the $\delta \mathscr{A}_\mu $ terms by a radial and time function times a $\theta$ and $\varphi$ function. We choose a basis of functions $Y(\theta,\varphi)$ that lives on the angular manifold and satisfies the angular Laplacian eigenvalue problem
\begin{equation}
    P^2 \frac{\partial^2Y}{\partial\theta^2}+(PP'+P^2\cot{\theta} )\frac{\partial Y}{\partial\theta}+\csc{\theta}\frac{\partial^2Y}{\partial\varphi^2}=-\lambda_{1} P Y,
    \label{maxwellangular}
\end{equation}
with eigenvalue $\lambda_1$. As a consequence we can write
\begin{equation}
    \delta \mathscr{A}^{P}_\mu = \left(a_0(t,r)Y(\theta, \phi),a_1(t,r)Y(\theta, \phi),a_2(t,r)Y_{,\theta}(\theta, \phi),a_2(t,r)Y_{,\phi}(\theta,\phi)\right)
\end{equation}
and 
\begin{equation}
    \delta \mathscr{A}^{A}_\mu = \left(0,0,-a_3(t,r)\frac{Y_{,\varphi}(\theta,\varphi)}{P(\theta)\sin{\theta}} , a_3(t,r)Y_{,\theta}(\theta,\varphi)P(\theta) \sin{\theta}\right),
\end{equation}
where the functions $a_i(t,r),i=0,1,2,3$ are introduced and the commas represent derivatives with respect to the coordinates following it. These perturbations neither transform evenly nor oddly under parity transformations, since the spacetime acceleration breaks this metric isometry.

The angular $\theta$ and $\varphi$ components in \eqref{maxwellangular} can be separated by introducing $Y(\theta,\varphi)= e^{im \varphi}\chi_1(\theta)$ and further simplified by introducing the angular tortoise $z$ defined in \eqref{angular tortoise} to obtain
\begin{equation}
    \frac{d^2\chi_1}{d z^2}-(m^2  -\,_1V_\theta)\chi_1 =0,
    \label{finalangularmaxwelleq}
\end{equation}
where
\begin{equation}
 _1V_\theta=\lambda_1 P(\theta) \sin^2{\theta}.
 \label{finalangularmaxwellpotential}
\end{equation}

We now turn our attention to the radial and time sectors. First, it is important to fix a gauge. We find that Maxwell's equation are invariant under the general gauge transformation 
\begin{equation}
    A_\mu \rightarrow A_\mu + \partial_\mu \eta,
\end{equation}
where $\eta$ is a general function. Separating it into a radial plus time function times an angular function with the ansatz $\eta=\eta(t,r)Y(\theta,\varphi)$, we find out how each component of $\delta \mathscr{A}_\mu$ changes under the gauge transformation:
\begin{equation}
    a_0 \to a_0+\dot \eta,\quad a_1 \to a_1+\eta',\quad  a_2\to a_2+\eta \quad \text{and} \quad a_3 \to a_3.
\end{equation}
The prime denotes derivative with respect to $r$ and the dot represents derivatives in $t$. This set of gauge transformations allows us to choose a gauge where $a_2=0$.
By further separating time from space, using $a_i(t,r)= e^{-i\omega t}\,u_i(r)$, where $i=0,1,2,3$, we are ready to obtain the master equations for the components of the electromagnetic tensor.

On the axial sector, the $\theta$ component of \eqref{MaxwellEQ}, gives the same equations of motion as its $\varphi$ component. Using the tortoise coordinate \eqref{tortoise}, they can be written as 
\begin{equation}
    -\frac{d^2u_3}{dx^2}+\left(\omega ^2 - \,_1V_r\right) \,u_3=0
\end{equation}
with
\begin{equation}
    _1V_r=  \frac{\lambda_1 f(r) }{r^2}.
    \label{finalradialmaxwellpotential}
\end{equation}

The polar sector is slightly more complicated. Its equations of motion are the following: evaluating them on the $t$ component of \eqref{MaxwellEQ} gives
\begin{equation}
    r(u_0''+i\omega u_1')+2(u_0'+i\omega u_1)=\frac{\lambda_1 u_0}{rf(r)}
    \label{t}
\end{equation}
On the $r$ component,
\begin{equation}
    r^2(u_0'+i \omega u_1)=i\frac{\lambda_1}{\omega}f(r) u_1
    \label{r}
\end{equation}
and lastly on the $\theta$ and $\varphi$ components, we get
\begin{equation}
    u_0=\frac{i}{\omega}f(r)(f(r)u_1)'.
    \label{angle}
\end{equation}
Out of the three equations \eqref{t}, \eqref{r} and \eqref{angle}, one is linearly independent of the other two. To obtain a master equation we define 
\begin{equation}
    \phi_1=r^2(u_0'+i\omega u_1).
\end{equation}
Therefore, equation \eqref{r} becomes
\begin{equation}
    \phi_1=i\frac{\lambda_1}{\omega}f(r) u_1.
    \label{u1phi}
\end{equation}
Notice that expression \eqref{t} can be written as 
\begin{equation}
    \frac{d\phi_1}{dr}=\lambda_1\frac{u_0}{f(r)}.
    \label{phir}
\end{equation}
We are looking for a Schr\"{o}dinger-like equation here as well, so we compute the second derivative of $\phi$ and find that
\begin{equation}
    \frac{d^2\phi_1}{dx^2}=\lambda_1f(r)u_0'.
    \label{phiu0}
\end{equation}
we now use the fact that 
\begin{equation}
    u_0'=\frac{\phi_1}{r^2}-i\omega u_1,
\end{equation}
together with \eqref{u1phi} to obtain
\begin{equation}
    u_0'=\frac{\phi_1}{r^2}-\frac{\omega^2\phi_1}{\lambda_1f(r)}.
\end{equation}
Finally, plugging this in \eqref{phiu0} gives us
\begin{equation}
    \frac{d^2\,\phi_1}{dx^2}+(\omega^2 -\,_1V_r)\,\phi_1=0,
    \label{finalradialmaxwelleq}
\end{equation}
where the potential $_1V_r$ is the same as defined in \eqref{finalradialmaxwellpotential}
\begin{equation}
    _1V_r=\lambda_1\frac{f(r)}{r^2}.
\end{equation}
Therefore, we have concluded that polar modes obey the same differential equations as axial modes. Consequently, when subject to boundary conditions, they will exhibit the same spectrum and hence isospectrality is manifest once more.

\subsection{On the boundary conditions}\label{On the boundary conditions}
%%%%%%%%%%%%%%%%%%%%%%%%%%%%%%%%%%%%%%%%%%%%%%%%%%%%%%%%%%%%%%%%%%%%%%%%%%%%%%%%%%%%%%%%%%%%%%%%%%%%%%%%%%%%
%%%%%%%%%%%%%%%%%%%%%%%%%%%%%%%%%%%%%%%%%%%%%%%%%%%%%%
%%%%%%%%%%%%%%%%%%%%%%%%%%%%%%%%%%%%%%%%%%%%%%%%%%%%%%%%%%%%%%%%%%%%%%%%%%%%%%%%%%%%%%%%%%%%%%%%%%%%%%%%%%%%
\noindent
To summarize, we have seen that all these fields obey a Schr\"{o}dinger-like equation in their radial and angular sectors. Namely, consider $\chi_s(\theta)$ as the field's angular function labeled by spin $s=0,1/2,1$; we have found that generically it obeys a differential equation of the form
\begin{equation}
    \frac{d^2\chi_{s}}{dz^2}-(m^2-\,_{s}V_\theta)\chi_{s}=0,
    \label{general angular equation}
\end{equation}
where $_sV_\theta$ is given by \eqref{finalangularscalarpotential}, \eqref{finalangulardiracpotential} and \eqref{finalangularmaxwellpotential} for the scalar, Dirac and Maxwell fields respectively.

Consider now the radial function $\phi_s(r)$, labeled by spin $s=0,1/2,1$. We have that it follows the differential equation
\begin{equation}
    \frac{d^2\phi_s}{dx^2}+(\omega^2-\,_sV_r)\phi_s=0,
    \label{general radial equation}
\end{equation}
where the potential is given by \eqref{finalradialscalarpotential}, \eqref{finalradialdirapotential} and \eqref{finalradialmaxwellpotential} for the scalar, Dirac and Maxwell fields respectively.

One of the terms that was present in the separation of variables is $e^{im \varphi}$. The set of values $m$ can take depends on the period of the variable $\varphi$. Generically, we take it to be $2\pi C$, where $C$ is chosen such that one of the conical singularities can be removed. Therefore after a full $2\pi C$ turn the field must return to its original value, so we impose
\begin{equation}
    e^{im2 \pi C}=1\implies m=\frac{\hat m}{C},\quad \hat m=0,1,2,\dots.
    \label{m values}
\end{equation}
The exact value of $C$ will not be a major concern on this work; however, the interested reader can find the corresponding values for each choice of regularized rotation axis in \cite{Silva:2026tev}.

Additionally, we should impose boundary conditions on the radial and angular functions. These will differ depending on the kind of problem we wish to study.

On the study of linear stability, to be performed in section \ref{Linear stability}, we are interested in the well posedness of the equations of motion. Therefore, we will impose that the radial and angular fields are square integrable, which essentially means that the fields vanish when the corresponding tortoise coordinates go to infinity. This statement will be more explicit in \ref{Linear stability}.

If we are interested on mode stability (the focus of section \ref{Mode stability}) or on the study of quasinormal mode, we must impose a specific kind of boundary conditions, which we shall discuss now. Consider the radial equation of motion and remember that the coordinate $r$ takes values between the event and cosmological horizons, which we shall label as $r\in (r_H,r_c)$. Independently of the spin perturbation, we find that as $r\to r_H$ or $r\to r_c$, the potential $_sV_r$ goes to zero. As a consequence of equation \eqref{general radial equation}, 
\begin{equation}
    \phi_s \sim \mathcal{A}e^{i \omega x}+\mathcal{B}e^{-i\omega x}\quad (r\to r_H \text{ or } r\to r_c),
\end{equation}
where $\mathcal{A}$ and $\mathcal{B}$ are integration constants.
Quasinormal modes are defined as ingoing waves at the event horizon and outgoing waves at the acceleration horizon.
Consequently we choose
\begin{equation}
    \phi_s\sim \begin{cases}
        e^{-i\omega x},\quad r\to r_H \quad (x\to -\infty), \\
        e^{i\omega x},\quad r\to r_c \quad  (x\to +\infty).
    \end{cases}
    \label{BCQNM}
\end{equation}

The boundary conditions prescribed by \eqref{BCQNM} are actually responsible for much more than what a first glance might tell. They make the eigenvalue equation \eqref{general radial equation} define a non hermitian problem which essentially comes back to influence the $\omega$ spectrum. Consequently, the eigenvalues $\omega^2$ do not necessarily need to be real. In fact, $\omega$ is usually complex valued:
\begin{equation}
    \omega=\omega_R+i\omega_I.
\end{equation}
Its physical meaning is that quasinormal modes define oscillations with time varying amplitudes. While $\omega_R$ dictates the frequency of the oscillation, $\omega_I$ gives us the mode's damping (if $\omega_I<0$) or it tells us about its instability timescale (if $\omega_I>0$). This is easily seen by the following equality
\begin{equation}
    e^{-i\omega t}=e^{-i\omega_Rt}e^{\omega_It}.
\end{equation}
If we wish for a stable black hole, we must have $\omega_I<0$. This problem will be addressed in the mode stability section \ref{Mode stability}.

Consider now the angular part. The range of values that $\theta$ takes differs fundamentally depending on the accelerating regime we are considering. On the regular regime $\theta \in (0,\pi)$; on the first super accelerating regime $\theta \in (\theta_+,\pi)$ and for the second super accelerating regime we either have $\theta\in (0,\theta_-)$ or $\theta\in (\theta_+,\pi)$. To proceed in a generic way, let us say that $\theta\in(\theta_H,\theta_c)$, where $\theta_H$ and $\theta_c$ are clearly defined for each accelerating regime in question.

We note that the potential $_sV_\theta$ goes to zero at $\theta_H$ and $\theta_c$ for $s=0$ and $s=1$. However, it becomes some constant for $s=1/2$; let us call it $_{1/2}V^\theta$. Consequently, the asymptotics of \eqref{general angular equation} at the boundaries of $\theta$ are
\begin{equation}
    \chi_s=\mathcal{E}e^{\sqrt{m^2-\,_{s}V^\theta}z}+\mathcal{F}e^{-\sqrt{m^2-\,_{s}V^\theta}z} \quad (\theta \to \theta_H \text{ or } \theta\to\theta_c),
\end{equation}
where $\,_sV^\theta=0$ for $s=0,1$ and $\mathcal{E},\mathcal{F}$ are separation constants.
The relevant solution to consider is the one regular at the boundary
\begin{equation}
    \chi_s \sim \begin{cases}
        e^{\sqrt{m^2-\,_{1/2}V^\theta}z},\quad \theta\to \theta_H \quad (z\to -\infty), \\
        e^{-\sqrt{m^2-\,_{1/2}V^\theta}z}, \quad \theta\to \theta_c \quad (z\to +\infty).
        \label{angularboundarycondition}
    \end{cases}
\end{equation}
This allows us to conclude that even though the boundary of $\theta$ can change, due to the introduction of the new axis of symmetry, the separation constant will always be a real number. This is so as the behavior of $\chi_s$ at the poles is not oscillatory but exponentially decaying, allowing for the formulation of a hermitian problem. Additionally, given the way we are treating these fields, the conical singularities of the black hole (present on the boundaries of the domain of $\theta$) do not lead to irregularities of $\chi_s$. However, they do influence the field $\chi_s$ once they are regularized. This is through their effect on the period of $\varphi$, thus changing the parameter $m$ as prescribed by \eqref{m values} and then finally affecting $\chi_s$ through equation \eqref{general angular equation}.

Finally, there is also the possibility of studying the transmission and reflection scattering problem, which is closely related with the study of greybody factors. In that case, the boundary conditions for the angular equation remain, but for the radial equation change.
To describe Hawking radiation, we consider that radiation comes out of the black hole. Some of it gets reflected there while another part is transmitted away. This leads us to impose
\begin{equation}
    \phi_s \sim \begin{cases}
        e^{i\omega x}+R(\omega)e^{-i\omega x },\quad r\to r_H \quad (x\rightarrow -\infty )
         \\
         T(\omega)e^{i\omega x},\quad r\to r_c \quad (x\rightarrow +\infty)
         \label{Transmissionconditions}
    \end{cases}
\end{equation}
where $R(\omega)$, $T(\omega)$ are the reflection and transmission coefficients respectively. The greybody factor is therefore defined as the probability of radiation being transmitted
\begin{equation}
    \Gamma(\omega)=|T(\omega)|^2.
\end{equation}

Let us further notice how $R(\omega)$ and $T(\omega)$ are related. Notice that a mode of frequency $\omega$ solves the master equation with \eqref{Transmissionconditions}, then a mode with frequency $-\omega$ solves the same equation but with
\begin{equation}
    \phi_s \sim \begin{cases}
                e^{-i\omega x}+R(-\omega)e^{i\omega x },\quad r\to r_H \quad (x\rightarrow -\infty ),
         \\
         T(-\omega)e^{-i\omega x},\quad r\to r_c \quad (x\rightarrow +\infty).
    \end{cases} 
\end{equation}
It is possible to construct a conserved current of \eqref{general radial equation}. It is given by
\begin{equation}
    J_s=\frac{1}{2i}\left(\phi_s(-\omega) \frac{d\phi_s(\omega)}{dx}-\phi_s(\omega) \frac{d\phi_s(-\omega)}{dx} \right).
\end{equation}
Evaluating it when $x$ goes $+\infty$ and equating that when $x\to -\infty$ allows us to conclude that 
\begin{equation}
    R(\omega) R(-\omega) +T(\omega) T(-\omega) =1.
    \label{GBFconservation}
\end{equation}
This can be interpreted as a conservation of energy condition. To see why, suppose $\omega \in \mathbb{R}$, then $\phi(-\omega)=\phi^*(\omega)$. Consequently  $T(-\omega)=T^* (\omega)$ and $R(-\omega)=R^*(\omega)$; therefore \eqref{GBFconservation} tells us that $|R|^2+|T|^2=1$. If there was only one horizon from which Hawking radiation could originate from, this analysis would be enough. However, Hawking radiation can emerge from the cosmological horizon $r_c$ too. As such, we should consider the scattering problem for this one as well. In practice, we will see that the problems can be considered independently and the transmission coefficient will be the same.

Let us denote $\phi_s'(\omega)$ as the field that satisfies this second scattering problem. The boundary conditions must be
\begin{equation}
    \phi_s' \sim \begin{cases}
        T'(\omega)e^{-i\omega x}, \quad r\to r_H\quad (x\rightarrow -\infty),
         \\
         e^{-i\omega x}+R'(\omega)e^{i\omega x }, \quad r\to r_c\quad (x\rightarrow +\infty).
         \end{cases}
\end{equation}

Since the master equation is a second order differential equation, all of its solutions can be written as a linear combination of $\phi(\omega)$ and $\phi(-\omega)$. Indeed, we can write 
\begin{equation}
    \phi'(\omega)=-\frac{R(-\omega)}{T(\omega)}\phi(\omega)+\frac{1}{T(-\omega)}\phi(-\omega),
\end{equation}
where we deduce that 
\begin{equation}
    R'(\omega)=-\frac{T(\omega)}{T(-\omega)}R(-\omega) \quad \text{and} \quad T'(\omega)= T(\omega).
    \label{scatteringout}
\end{equation}
So we conclude that the transmission coefficients are actually the same and therefore so should the greybody factors. By swapping $\omega$ with $-\omega$ we obtain the expressions for $R'(-\omega)$ and $T'(-\omega)$ which allow us to find that $R'(\omega)R'(-\omega)+T'(\omega)T'(-\omega)=1$.

\section{Linear stability} \label{Linear stability}
%%%%%%%%%%%%%%%%%%%%%%%%%%%%%%%%%%%%%%%%%%%%%%%%%%%%%%%%%%%%%%%%%%%%%%
%%%%%%%%%%%%%%%%%%%%%%%%%%%%%%%%%%%%%%%%%%%%%%%%%%%%%%%%%%%%%%%%%%%%%%
\noindent
In this section, we aim to prove that the master equation \eqref{general radial equation} defines a positive, self adjoint operator and therefore there is no mode with $\omega^2\leq0$, implying that $\omega$ is a real number and consequently that there are no instabilities. As a requirement, we impose that $\phi_s\in L^2$ is squared integrable. 

The following proof relies on building an energy integral from \eqref{general radial equation}. This can be obtained by multiplying the field complex conjugate $\phi_s^*$, on both sides of the master equation and integrating by parts. Since the boundary terms vanish (given that $\phi_s\in L^2$), we find that
\begin{equation}
    \int_{-\infty}^{+\infty} \left|\frac{d\phi_s}{dx} \right|^2+\,_sV_r |\phi_s |^2 dx = \omega^2 \int_{-\infty}^{+\infty} |\phi_s|^2dx.
    \label{openergy}
\end{equation}
It straightforwardly follows that if $_sV_r\geq0$, then $\omega^2\geq0$. Our aim, therefore, shall be precisely to prove that the potential is positive definite. Remember however, that $_sV_r$ depends on the constant $\lambda_s$ and therefore bounds on this constant must be obtained. This can be done so by constructing a similar energy for the angular equation in an analogous way
\begin{equation}
        \int_{-\infty}^{+\infty} \left|\frac{d\chi_s}{dz}\right|^2 +m^2|\chi_s|^2 dz= \int_{-\infty}^{\infty} \,_sV_\theta|\chi_s|^2 dz,
    \label{angularenergy}
\end{equation}
where $\lambda_s$ is present inside of $_sV_\theta$. On the following subsections, we will consider each accelerating regime and spin field separately.

\subsection{Maxwell Field}
\noindent
The Maxwell field is the easiest field to prove linear stability. Its potential is given by \eqref{finalradialmaxwellpotential}
\begin{equation}
    _1V_r=\lambda_1\frac{f(r)}{r^2}.
\end{equation}
Since the function $f(r)\geq0$ for the region under study, we find that in order that $_1V_r\geq0$, it suffices to show that $\lambda_1\geq0$.

We look at the angular potential given by \eqref{finalangularmaxwellpotential}
\begin{equation}
    _1V_\theta=\lambda_1P(\theta)\sin^2\theta.
\end{equation}
Plugging this in \eqref{angularenergy} we obtain that
\begin{equation}
    \lambda_1=\frac{\int_{-\infty}^{+\infty} \left(\left|\frac{d\chi_1}{dz}\right|^2 +m^2|\chi_1|^2 \right)dz}{\int_{-\infty}^{+\infty}P(\theta)\sin^2(\theta)|\chi_1|^2dz}\geq0
    \label{Maxwellconstantinequality}
\end{equation}
This inequality is obtained since $P(\theta)$ is necessarily positive for the range of $\theta$ under consideration, as well due to $m$ being necessarily a real quantity as discussed in \ref{On the boundary conditions}, more specifically given by \eqref{m values}. Hence, we have shown that $_1V_r\geq0$ and therefore $\omega^2\geq0$ for the Maxwell field. As we have seen, the argument above does not depend on the accelerating regime under question and therefore generalizes to any one.

\subsection{Dirac Field}
\noindent
The Dirac field is governed by the partner potentials $_{1/2}V_r^\pm$ which are related by supersymetry \eqref{finalradialdirapotential}
\begin{equation}
    _{1/2}V_r^\pm=\mathscr{W}^2\pm \frac{d\mathscr{W}}{dx}.
    \label{superpotential2}
\end{equation}
For these kind of fields, we will not need to look at the angular equation of motion. Instead, it is useful to introduce the $S$-deformability method. The main idea is to define the operator 
\begin{equation}
    D=\frac{d}{dx}+S,
\end{equation}
where $S$ is some function of $r$. As a consequence, one can rewrite \eqref{openergy} as
\begin{equation}
    \int_{-\infty}^{+\infty} \left|D\phi_s\right|^2+\,_s\tilde V_r|\phi_s|^2dx=\omega^2\int_{-\infty}^{+\infty}  |\phi_s|^2dx,
    \label{Sopenergy}
\end{equation}
where
\begin{equation}
    _s\tilde{V}_r=\,_sV_r+ \frac{dS}{dx}-S^2.
    \label{S deformation transformation}
\end{equation}
Therefore, instead of proving $_sV_r\geq0$, one can alternatively choose an appropriate $S$ such that it becomes much easier to prove $_s\tilde V_r\geq0$. This is specially useful for the Dirac fields, because their potential \eqref{superpotential2} comes from a super potential $\mathscr{W}$. As a result, there is a very simple choice for the function $S$ for each $\pm$ super potential. It is
\begin{equation}
    S^\pm=\mp\mathscr{W}.
    \label{S deformation dirac}
\end{equation}
Using $S$ from \eqref{S deformation dirac} in equation \eqref{S deformation transformation}, implies that the transformed potential vanishes:
\begin{equation}
     _{1/2}V_r^\pm \to \,_{1/2}\tilde V_r^\pm=0.
\end{equation}
Finally, it trivially follows from \eqref{Sopenergy} that $\omega^2\geq0$, thus concluding our proof that Dirac perturbations are stable. The argument performed, once more, does not depend on the accelerating regime under question, so it generalizes easily to any of them. 

\subsection{Scalar field}
\noindent
The scalar field is governed by the potential \eqref{finalradialscalarpotential}
\begin{equation}
    _0V_r=f(r)\left(\frac{\tilde\lambda_0}{r^2}+\frac{2M}{r^3}-\frac{2Q^2}{r^4} \right),
\end{equation}
where $\tilde \lambda_0=\lambda_0-1/3+A^2Q^2/3$.
Given that we wish to prove $_0V_r\geq0$, we end up having to prove that
\begin{equation}
    \tilde \lambda_0\geq-\frac{2M}{r}+\frac{2Q^2}{r^2},\quad \text{for all } r\in (r_H,r_c).
    \label{lambdaconstraint}
\end{equation}

For simplicity, first consider that the black hole is uncharged ($Q=0$). The constraint \eqref{lambdaconstraint} can be expressed as 
\begin{equation}
    \tilde \lambda_0\geq \frac{-2M}{r_c}=\begin{cases}
        -2MA, \quad r_+<r_A, \\
        -1,\quad r_A<r_+.
    \end{cases}
    \label{lambdaconstraintuncharged}
\end{equation}
We shall now see whether such bound really is satisfied. That leads us to look at the angular equation whose angular potential is \eqref{finalangularscalarpotential}
\begin{equation}
    _0V_\theta =P(\theta)\left( \tilde\lambda_0\sin^2\theta+2MA\sin^2\theta\cos\theta-2A^2Q^2\sin^2\theta\cos^2\theta\right).
\end{equation}
For zero charge, equation \eqref{angularenergy} tells us that
\begin{equation}
\begin{aligned}
    &\tilde \lambda_0=\frac{\int_{-\infty}^{+\infty} \left(\left|\frac{d\chi_0}{dz}\right|^2 +|\chi_0|^2 \left( m^2-2MAP(\theta)\sin^2\theta\cos\theta\right)\right)dz}{\int_{-\infty}^{+\infty} \left(P(\theta)\sin^2\theta |\chi_0|^2\right)dz}\geq
    \\&  \geq -2MA\frac{\int_{-\infty}^{+\infty} P(\theta) \sin^2 \theta \cos \theta |\chi_0|^2dz}{P(\theta)\sin^2\theta |\chi_0|^2dz}\geq -2MA\max_\theta [\cos \theta],
    \label{unchargedlambdaconstrainangular}
\end{aligned}
\end{equation}
where $\max$ denotes the maximum function.
Since the range of $\theta$ allowed depends on the accelerating regime, we find that while on the regular accelerating regime, $\max [\cos \theta] =1$, on the super accelerating regime $\max[\cos\theta]=r_A/r_+$. Consequently, 
\begin{equation}
    \tilde \lambda_0 \geq \begin{cases}
        -2MA \quad r_+<r_A\\
        -1\quad r_A<r_+
    \end{cases}.
\end{equation}
This is exactly what we sought in \eqref{lambdaconstraintuncharged}. As such, for the uncharged black hole we have that $\omega^2\geq0$ and thus scalar fields are linearly stable on both the C-metric regular and super accelerating regimes.

On the other hand, if charge is present, the problem gets more complicated. In fact, we do not find linear stability on the hole range of parameters. Indeed, to find $\tilde\lambda_0$ satisfying \eqref{lambdaconstraint}, we should look at the maximum of
\begin{equation}
    g(r)\coloneq -\frac{2M}{r}+\frac{2Q^2}{r^2}.
    \label{g(r)}
\end{equation}
This function does not have a local maximum, but instead a minimum present at $r=2Q^2/M$, therefore its maximum is obtained at the boundaries. For the regular accelerating regime, the maximum is obtained at $r=r_A$, giving $\max g(r)=-2MA+2Q^2A^2$. On the initial parameter range of the first super accelerating regime\footnote{This corresponds to the first acceleration values obtained right after entering the super accelerating regime from the regular one.}, the maximum is instead reached when $r=r_+$. This holds for as long as $(-2Mr_A+2Q^2)/r_A^2<(-2Mr_++2Q^2)/r_+^2$. When we remain on the first super accelerating regime, but instead have $(-2Mr_++2Q^2)/r_+^2<(-2Mr_A+2Q^2)/r_A^2$, we find that the maximum of $g(r)$ is obtained for $r=r_A$. Afterwards, when the second super accelerating regime is reached, the maximum is instead found when $r=r_-$. Summarizing, we need to find $\tilde \lambda_0$ such that
\begin{equation}
    \tilde\lambda_0\geq \begin{cases}
    -2MA+2Q^2A^2, \quad r_-<r_+<r_A, \\
    (-2Mr_+ +2Q^2)/r_+^2, \quad r_-<r_A<r<r_+  \quad \wedge \quad (-2Mr_A+2Q^2)/r_A^2<(-2Mr_++2Q^2)/r_+^2, %\Leftrightarrow r_+<\frac{M}{1-MA+Q^2A^2}
    \\ 
     -2MA+2Q^2A^2, \quad r_-<r_A<r<r_+ \quad \wedge \quad (-2Mr_A+2Q^2)/r_A^2>(-2Mr_++2Q^2)/r_+^2, %\Leftrightarrow r_+>\frac{M}{1-MA+Q^2A^2}
     \\
    (-2Mr_- +2Q^2)/r_-^2, \quad r_A<r_-<r<r_+.
    \end{cases}
    \label{lambda condition charged radial}
\end{equation}

To discover to what extend these conditions hold we should look at the angular equation again \eqref{angularenergy}. Analogously to the uncharged version, \eqref{unchargedlambdaconstrainangular}, this sets that
\begin{equation}
    \tilde \lambda_0\geq \min_\theta [-2MA \cos \theta +2A^2Q^2\cos^2\theta].
\end{equation}
Therefore, we ought to search for the minimum of the function 
\begin{equation}
    h(\theta)\coloneq-2MA \cos\theta+2A^2Q^2\cos^2\theta.
\end{equation} 
Indeed, there is quite the resemblance of $h(\theta)$ with $g(r)$ in \eqref{g(r)}, but we shall see that appearances can be deceiving.
For the regular accelerating regime, we find that the minimum is obtained when $\theta=0\implies h(0)=-2MA+2A^2Q^2$. On the first super accelerating regime the minimum is given at $\theta=\theta_+=\arccos(r_A/r_+)\implies h(\theta_+)=\left(-2Mr_++2Q^2\right)/r_+^2$. By further increasing acceleration, we come across two possibilities. Either $\theta \in (0,\theta_-)$, with $\theta_-=\arccos (r_A/r_-)$ or $\theta \in (\theta_+,\pi)$. The last option actually gives the same minimum $h(\theta_+)$. The first option attains the minimum at $\theta_-$, where $h(\theta_-)=\left(-2Mr_-+2A^2Q^2\right)/r_-^2$. Summarizing, we have found that for the regular and first super accelerating regimes
\begin{equation}
    \tilde \lambda_0\geq\begin{cases}
        -2MA+2A^2Q^2, \quad r_-<r_+<r_A\\
        (-2Mr_++2Q^2)/r_+^2,\quad r_-<r_A<r_+
    \end{cases}
\end{equation}
while for the second super accelerating regime that either 
\begin{equation}
    \tilde\lambda_0 \geq (-2Mr_++2Q^2)/r_+^2, \text{ if } \theta \in (\theta_+,\pi) \quad \text{or} \quad \tilde \lambda_0 \geq (-2Mr_-+2Q^2)/r_-^2, \text{ if } \theta \in (\theta_-,\theta_+).
\end{equation}

Hence, we conclude that we have a positive definite potential $_0V_r$ in the regular accelerating regime. For the first super accelerating regimes, this also holds as long as 
\begin{equation}
(-2Mr_A+2Q^2)/r_A^2<(-2Mr_++2Q^2)/r_+^2.
\end{equation}

On the other hand, on the second super accelerating regime, we only find that $_0V_r$ becomes positive definite if we consider that $\theta \in (0,\theta_-)$. Therefore, only on these specific conditions do we find that $\omega^2\geq0$ for scalar fields.

Given that, we have not been able to prove linear stability throughout the whole parameter space, it is worth taking into consideration the possibility to use the $S$ deformability method here. Indeed, the ansatz $S=-f(r)/r$, seems quite attractive as the resulting potential gives
\begin{equation}
    _0\tilde V_r=f(r)\left( \frac{\tilde \lambda_0}{r^2}-\frac{2MA^2}{r}+2A^2\right)
\end{equation}
To impose positive definiteness on $_0\tilde V_r$, we would require that
\begin{equation}
    \tilde \lambda_0\geq 2A(Mr-r^2),\quad \text{for all } r\in(r_H,r_c).
\end{equation}
Unfortunately, we have not been able to obtain such bounds on $\tilde \lambda_0$ from the angular equation. Ideally, one would consider using the $S$ deformability method on the angular equation as well. However, this has proved to be not very successful as it is difficult to find the ideal $S$ that could lead to such inequality.

\section{Mode stability} \label{Mode stability}
%%%%%%%%%%%%%%%%%%%%%%%%%%%%%%%%%%%%%
%%%%%%%%%%%%%%%%%%%%%%%%%%%%%%%%%%%%%%%%%
\noindent
We now come back to quasinormal modes, which are defined by the boundary conditions that the radial function $\phi_s$ satisfies \eqref{BCQNM}. As we have discussed in subsection \ref{On the boundary conditions}, a mode with $\omega_I>0$ is deemed unstable, because as time advances, it grows exponentially. Up to some point, the field is so large that back reaction effects on the curvature must be taken into account. 

The objective of this section is to review a known proof regarding the stability of quasinormal modes. Indeed we shall find that it is not possible for $\omega_I>0$ and therefore that there are no such instabilities.

To start off with, we build a similar energy as \eqref{openergy}, but now boundary terms do not vanish, instead we obtain
\begin{equation}
    \int_{-\infty}^{+\infty} \left| \frac{d\phi_s}{dx}\right|^2 +\,_sV_r|\phi_s|^2 \,dx= \omega^2\int_{-\infty}^{+\infty}|\phi_s|^2 \,dx + i\omega (|\phi_s(+\infty)|^2+|\phi_s(-\infty)|^2).
    \label{energy}
\end{equation}
From now on, allow us to omit the spin label $s$ since the procedure is generic. Let us decompose the frequency into its imaginary and real parts in \eqref{energy}. Because the left hand side of \eqref{energy} is a real number for any kind of perturbation, the imaginary part of the right hand side has to be zero. The conditions is
\begin{equation}
    2\omega_R\omega_I \int_{-\infty}^{+\infty}|\phi|^2 dx +\omega_R (|\phi(+\infty)|^2+|\phi(-\infty)|^2)=0.
    \label{I}
\end{equation}
To proceed, we have to consider that $\omega_R\neq0$, allowing us to divide both sides by $\omega_R$. Then, solving for $\omega_I$, we get
\begin{equation}
    \omega_I=-\frac{|\phi(+\infty)|^2+|\phi(-\infty)|^2}{2\int_{-\infty}^{+\infty}|\phi|^2 dx}.
\end{equation}
This clearly shows that the right hand side is not positive. Thus, we conclude that $\omega_I\leq0$ and the fields are stable.

Now we have to deal with the $\omega_R=0$ case. Even though a mode with $\omega_R=0$ is not really oscillating, we can show that $\omega_I$ has to be negative as well. To see why, we look at the real part of \eqref{energy}. Suppose that $\omega_R=0$; then equation \eqref{energy} becomes
\begin{equation}
    \int_{-\infty}^{+\infty} \left| \frac{d\phi}{dx}\right|^2 +\,_sV_r|\phi|^2 \,dx= -\omega_I^2\int_{-\infty}^{+\infty}|\phi|^2 \,dx - \omega_I (|\phi(+\infty)|^2+|\phi(-\infty)|^2).
    \label{wr=0}
\end{equation}
From the linear stability section, we have found that $_sV_r$ is positive definite expect for a given parameter region for the scalar field, which we will ignore here. This leads us to conclude that the left hand side of \eqref{wr=0} is greater or equal to zero; therefore we again obtain stability, $\omega_I\leq0$.

Finally, consider the case $\omega_R\neq0$ and $\omega_I=0$. We shall see that this is also not allowed. Using \eqref{I} we conclude that this hypothesis imposes that
\begin{equation}
    |\phi(+\infty)|^2=-|\phi(-\infty)|^2
\end{equation}
and therefore the field $\phi(\pm\infty)=0$. By using lemma $4.1$ from \cite{TeixeiradaCosta:2019skg}, we deduce that $\phi$ vanishes identically in all $r$ domain. Therefore, there are no non trivial purely oscillating modes, also known as normal modes.

The arguments performed here are completely general and do not depend on the spacetime regime under question. Thus we conclude that both the regular and super accelerating regimes are modally stable.

\section{Quasinormal mode families and greybody factors}\label{Quasinormal mode families}
%%%%%%%%%%%%%%%%%%%%%%%%%%%%%%%%%%%%%%%%%%%%%%%%%%%%%%%%%%%%%%%%%%%%%%
%%%%%%%%%%%%%%%%%%%%%%%%%%%%%%%%%%%%%%%%%%%%%%%%%%%%%%%%%%%%%%%%%%%%%%
\noindent
The quasinormal modes of accelerating black holes have been the subject of study of many works \cite{Destounis:2022rpk, Lei:2023mqx, Destounis:2020pjk, BarraganAmado:2023wxt, Chen:2024rov}. It has been found that they can be classified into various families, namely: the photon surface (or WKB) modes, the purely accelerating modes and the near extremal modes. 
This classification is based on the properties that these fields are characterized by. 

Photon surface modes are localized on the photon surface of the black hole and it was established that their real frequency corresponds to the angular frequency of photons spiraling around the black hole on circular null geodesics, whereas their imaginary part corresponds to the associated Lyapunov exponent of the geodesic \cite{Moura:2026ken, Cardoso:2008bp}.

Purely accelerating modes are present when the mass and charge parameters are close to zero, $M,Q\to 0$, but acceleration is non vanishing $A\neq0$. These are purely damped and localized on the acceleration horizon. An analytical computation of these modes has been performed in \cite{Zhou:2024axx} and they are found to resemble the QNM of pure de Sitter spacetimes \cite{Lopez-Ortega:2012xvr, A. Vasy}.

Near extremal mode families are present when the black hole is close to extermality. On these cases, the QNMs localize on the near extremal horizon of the black hole. On the regular accelerating regime of the C-metric, there are two such families corresponding to each extremal limit that is taken; namely, the accelerating Bertotti-Robinson limit where $r_-\sim r_+<r_A$ and the accelerating Nariai limit where $r_-<r_+\sim r_A$. While the Bertotti-Robinson family modes are purely damped, the Nariai modes have a non vanishing real oscillation frequency.

It is important to note that, while near extremal and purely accelerating families are present only for specific regions of the black hole parameters, the photon surface modes exist for general values $M,Q,A$. However, these last ones are deformed into the other families as the black hole parameter reach their regime of applicability.

In this section, our main focus lies on how this classification broadens to the super accelerating regime of the black hole. Indeed, we expect that the new families are easily obtained. In fact, there should be a super accelerating photon surface mode family, and new near extremal families related with the new extremal limits possible on this regime. Namely, the super accelerating Bertotti-Robinson ($r_A\sim r_-<r_+$) and the super accelerating Nariai ($r_A<r_-\sim r_+$) configurations.

Additionally, the study of greybody factors can be an interesting subject. In most cases, when the QNMs are analytically computable, so is the GBF. In the following subsections, we derive and review some analytical results regarding the QNM spectra and the GBF of accelerating black holes while additionally obtaining new results for the super accelerating regime as well spin $1/2$ and $1$ fields.

\subsection{Formalism}
\label{Formalism}
\noindent
With the exception of the photon surface mode family which will be discussed in \eqref{photon surface modes}, we will see that the resulting potential for the radial equation of motion can be approximately described by the P\"{o}schl Teller potential or its modified version, respectively given by
\begin{equation}
    _sV_r=\frac{_sV_0}{\cosh^2 (k_Hx)},\quad  \text{or}\quad  _sV_r=\frac{_s\tilde V_0}{\sinh^2(k_Hx)}.
    \label{THEPOTENTIALS}
\end{equation}
We will show that $k_H$ will represent the surface gravity of some horizon, $x$ is the typical tortoise coordinate \eqref{tortoise} and $_sV_0,\,_s\tilde V_0$ are constants to be found. One should be careful with the modified version of the P\"{o}schl Teller potential, as it diverges at $x\to0$; therefore we are restricted to impose that the field $\phi\to 0$ there. Additionally, it is worth pointing out that one of such potentials can be mapped onto another by the transformation \begin{equation}
    \cosh^2(k_Hx)=-\sinh^2(k_Hx-i\pi/2)=-\sinh^2(k_H \tilde x).
    \label{complextranslation}
\end{equation}
Since a translation $x\to \tilde x=x-i\pi/(2k_H)$ leaves the kinetic term $d^2/dx^2$ invariant, working with the P\"{o}schl Teller potential is thus equivalent to working with its modified version as long one swaps the sign of $_sV_0\to\, _s\tilde V_0=-\,_sV_0$.

Focusing our attention to the non modified P\"{o}schl Teller potential. The QNM spectrum and GBF are quite well known here. If $_sV_0/k_H^2>1/4$, then the QNM are by \cite{Ferrari:1984zz} 
\begin{equation}
    \omega=k_H \left(\pm\sqrt{\frac{V_0}{k_H^2}-\frac{1}{4}} -i\left(n+\frac{1}{2}\right)\right),\quad n=0,1,2\dots.
    \label{teller}
\end{equation}
and the greybody factors \cite{Ferrari:1984zz, Landau:1991wop} 
\begin{equation}
\Gamma(\omega)=|T(\omega)|^2=\frac{\sinh^2\left(\frac{\pi\omega}{k_H}\right)}{\sinh^2\left(\frac{\pi\omega}{k_H}\right)+\cosh^2\left[\pi\sqrt{ \frac{_sV_0}{k_H^2}-\frac{1}{4}} \right]}.
\label{Tteller}
\end{equation}
On the other hand, if $_sV_0/k_H^2<1/4$, we obtain purely damped modes
\begin{equation}
    \omega=-i|k_H| \left( n+\frac{1}{2}+\sqrt{-\frac{V_0}{k_H^2}+\frac{1}{4}}\right),\quad n=0,1,2\dots.
    \label{wPo2}
\end{equation}
and the GBF become
\begin{equation}
\Gamma(\omega)=|T(\omega)|^2=\frac{\sinh^2\left(\frac{\pi\omega}{k_H}\right)}{\sinh^2\left(\frac{\pi\omega}{k_H}\right)+\cos^2\left[\pi\sqrt{ -\frac{_sV_0}{k_H^2}+\frac{1}{4}} \right]}.
\label{TPo2}
\end{equation}
Otherwise, considering the modified P\"{o}schl Teller potential, the condition $_s\tilde V_0/k_H^2>-1/4$ will lead to \eqref{wPo2} and \eqref{TPo2}, but with $_s\tilde V_0= -_sV_0$; whereas the condition $_s\tilde V_0/k_H^2<-1/4$ gives \eqref{teller} and \eqref{Tteller} again with $_s\tilde V_0= -_sV_0$. For a more detailed discussion on the QNM problem in the P\"{o}schl Teller potentials we reference the reader to \cite{Cardona:2017scd}.

In the next subsections, with exception of \ref{photon surface modes}, we will be obtaining these potentials and computing $_sV_0$ or $_s\tilde V_0$. However, the massless Dirac field works slightly different. In this case, the resulting potential will have a different form 
\begin{equation}
_{1/2}V_r^{\pm}=\frac{V_0\mp\sqrt{V_0}k_H\sinh{(k_Hx)}}{\cosh^2{(k_Hx)}} \quad \text{or}\quad 
_{1/2}V_r^{\pm}=\frac{\tilde V_0\mp\sqrt{\tilde V_0}k_H\cosh{(k_Hx)}}{\sinh^2{(k_Hx)}}.% + \mathcal{O}(1/\sinh^3(kx)),
\label{fermionpotentiaslnearextremal}
\end{equation}
For the first potential, the QNM are given by \cite{Churilova:2021nnc}
\begin{equation}
    \omega=\pm V_0-i|k_H|\left(n+\frac{1}{2}\right),\quad n=0,1,2,\dots,
    \label{fermionQNMfrequencies}
\end{equation}
where the $\pm$ in front of $_sV_0$ are not related with the index of $_sV_0^\pm$, since partner potentials give the same non zero energy spectrum of quasinormal modes, as we have argued in subsection \ref{The Dirac field}. For the second potential in \eqref{fermionpotentiaslnearextremal}, we can apply the transformation discussed in \eqref{complextranslation} to obtain the QNMs \eqref{fermionQNMfrequencies} with $\tilde V_0=-V_0$, essentially yielding the same spectrum.

Lamentably, to the best of our knowledge, the greybody factor for a massless spin $1/2$ field, in the P\"{o}schl Teller potential have not yet been computed. Yet, it shall not be the goal of this work to perform this novel computation.

\subsection{Photon surface modes}\label{photon surface modes}
%%%%%%%%%%%%%%%%%%%%%%%%%%%%%%%%%%%%%%%%%%%%%%%%%%%%%%%%%%%%%%%%%%%%%%
%%%%%%%%%%%%%%%%%%%%%%%%%%%%%%%%%%%%%%%%%%%%%%%%%%%%%%%%%%%%%%%%%%%%%%
\noindent
The photon surface modes are obtained by performing the eikonal limit together with the application of WKB methods. In \cite{Moura:2026ken} one can find a comprehensive explanation on how these modes have been computed for the C-metric. For this work's purposes, we briefly mention that the eikonal limit is defined by a large separation constant $\lambda_s\to\infty$ and it follows that the different fields now become isospectral. Therefore we shall omit the label $s$ on the separation constant and simply name it by $\lambda$ from now on\footnote{For the Dirac field, we relabel the separation constant as $\lambda_{1/2}^2\to \lambda$ in the eikonal limit, as it is not $\lambda_{1/2}$, but its square that becomes $\lambda$.}. The QNM frequency spectra and the GBF have been obtained in this limit in \cite{Moura:2026ken}. The first is given by 
\begin{equation}
        \omega_R= \sqrt{\lambda} \frac{\sqrt{(Q^2-2Mr_c+r_c^2)(1-A^2r_c^2)}}{r_c^2},
        \label{wkb1}
\end{equation}
\begin{equation}
    \frac{\omega_I}{n+\frac{1}{2}}=-\frac{\sqrt{(Q^2-2Mr_c+r_c^2)(1-
    A^2r_c^2)(-10Q^2+12Mr_c+3(-1+A^2Q^2)r_c^2-2A^2Mr_c^3)}}{r_c^3}, \quad n=0,1,2,\dots
    \label{wkb2}
\end{equation}
where $n$ is the overtone number and $r_c$ is defined by
\begin{equation}
\begin{aligned}
     r_c&=\frac{-1+A^2Q^2}{3A^2M}+\frac{2\sqrt{1+A^4Q^4+A^2(9M^2-2Q^2)}}{3A^2M} \times \nonumber \\ &\times \cos{\left(\frac{1}{3}\arccos{\left[\frac{-2+A^2(-27M^2(1+A^2Q^2)+2Q^2(3-3A^2Q^2+A^4Q^4))}{2(1+A^4Q^4+A^2(9M^2-2Q^2))^{3/2}}\right]}\right)}.
\end{aligned}
\end{equation}
The GBF is given by 
\begin{equation}
    \Gamma(\omega)=\frac{1}{1+e^{2\pi iK}},
\end{equation}
with
\begin{equation}
    K=-i \left(\sqrt{\lambda}-\frac{\omega^2r_c^4}{\sqrt{\lambda}(Q^2-2Mr_c+r_c^2)(1-A^2r_c^2)}\right) \frac{r_c}{2\sqrt{-10Q^2+12Mr_c+3(-1+A^2Q^2)r_c^2-2A^2Mr_c^3}}.
\end{equation}
Notice that except for the imaginary part of $\omega$, there is a $\lambda$ dependency on all of these quantities. This is problematic because generically we do not know which values the separation constant takes. In reference \cite{Moura:2026ken}, this was treated by considering small $A$ approximations where the constant $\lambda$ admits a known solution \cite{BarraganAmado:2023wxt}. On the super accelerating regime however, this approximation is not appropriate of course.

We will not go through all the details as it will not be necessary for section \eqref{Stability of the Cauchy horizon}. But we just briefly propose a possible solution to the above problem. The idea is to apply the WKB method for the angular function \eqref{general angular equation}. In the eikonal limit, the angular potential is spin independent (hence we omit its $s$ label) and given by
\begin{equation}
    V_\theta= \lambda P(\theta)\sin^2\theta.
\end{equation}
A WKB argument will tell us that 
\begin{equation}
    m^2=V_\theta(\theta_0)-\left(\ell +\frac{1}{2} \right)\sqrt{-2\left(\frac{d^2}{dz^2}V_\theta\right)_{\theta=\theta_0}},
    \label{WKBlambda}
\end{equation}
where $\ell$ is a positive integer defining the angular momentum number\footnote{Naturally, the eikonal limit $\lambda \to \infty$, should coincides with the limit of $\ell\to \infty$.} and $\theta_0$ is the maximum of $V_\theta$. Given that $m$ is already determined by \eqref{m values}, and that $\lambda$ is present inside $V_\theta$, the idea is to invert \eqref{WKBlambda} and thus obtain $\lambda$ as a function of $m$ and the black hole's parameters, allowing us to write the QNM frequencies and the GBF only as a function of these last quantities plus the overtone and angular momentum numbers.

However, as we will see in the following subsections, we will find an alternative way to compute the separation constant reliably, although valid only for specific spacetime regimes and not as generically as this method would assure us.

\subsection{Purely accelerating modes}\label{Purely accelerating modes}
%%%%%%%%%%%%%%%%%%%%%%%%%%%%%%%%%%%%%%%%%%%%%%%%%%%%%%%%%%%%%%%%%%%%%%
%%%%%%%%%%%%%%%%%%%%%%%%%%%%%%%%%%%%%%%%%%%%%%%%%%%%%%%%%%%%%%%%%%%%%%
\noindent
The purely accelerating regime of the C-metric is defined when there is no black hole, $M=Q=0$. The following calculations can be thought as the zeroth order approximation to 
\begin{equation}
    r_+\ll r_A.
\end{equation}
As a consequence of neglecting the charge and mass parameters of the black hole, we obtain that the metric functions
\begin{equation}
    f(r)=1-A^2r^2,\quad  P({\theta})=1,\quad \text{and}\quad \Omega(r,\theta)=1.
\end{equation}
The new form of $P(\theta)$ recovers spherical symmetry of the spacetime and now the the separation constant is given by $\lambda_0=\ell(\ell+1)+1/3$, $\lambda_1=\ell(\ell+1)$ and $\lambda_{1/2}=\ell+1/2$, where $\ell$ is the angular momentum number.% $\ell=|s|,|s|+1,|s|+2\dots$.

The potential of the wave equation becomes the same for scalar and electromagnetic fields
\begin{equation}
    \,_s V_r(r)=(1-A^2r^2) \frac{\ell(\ell+1)}{r^2},
\quad s=0,1,
\label{acceleratepotential}
\end{equation}
while for massless Dirac fields it is
\begin{equation}
    _{1/2} V_\pm=\left(\ell+\frac{1}{2}\right)^2\left( \frac{1-A^2r^2}{r^2}\right)\mp \left(\ell+\frac{1}{2}\right)\frac{\sqrt{1-A^2r^2}}{r^2}.
    \label{acceleratepotentialfermion}
\end{equation}

To write it in the P\"oschl Teller potential form, we notice that the tortoise coordinate is easily computed to be
\begin{equation}
    x=\frac{1}{A}\arctanh(Ar).
\end{equation}
By inverting the function $x(r)$, we get
\begin{equation}
    r(x)=\frac{1}{A}\tanh(Ax).
\end{equation}
As such, plugging this in the potential \eqref{acceleratepotential} and \eqref{acceleratepotentialfermion} gives the modified P\"oschl Teller potential 
\begin{equation}
    _0V_r=\,_1V_r=\frac{k_A^2\ell(\ell+1)}{\sinh^2(k_Ax)},\quad _{1/2}V_r^\pm=k_A^2\frac{(\ell+1/2)^2\mp (\ell+1/2)\cosh(k_Ax)}{\sinh^2(k_Ax)},
\end{equation}
where we have defined $k_A=-A\eqcolon -k_A^R$, representing the negative of the Rindler surface gravity. We therefore conclude that
\begin{equation}
    _s\tilde V_0=k_A^2\ell(\ell+1),\quad s=0,1/2,1.
\end{equation}
Therefore, the QNM frequencies are given by
\begin{equation}
    \omega=\begin{cases}
        -i|k_A| \left(\ell +n+1 \right), \quad s=0,1 \\
         |k_A|\left(\pm \ell (\ell+1) -i(n+\frac{1}{2})\right),\quad s=1/2
    \end{cases}.
    \label{omegapureaccelerate}
\end{equation}
which agrees with what was found in \cite{Zhou:2024axx}. We thus observe that scalar and electromagnetic perturbation become isospectral on this regime. For these two kinds of perturbations, the greybody factor is also spin independent, giving
\begin{equation}
\Gamma(\omega)=|T(\omega)|^2=\frac{\sinh^2\left(\frac{\pi\omega}{k_A}\right)}{\sinh^2\left(\frac{\pi\omega}{k_A}\right)+\cos^2\left[\pi\left( \ell+n+1\right) \right]}.
\end{equation}

Actually, given the transformation $A\to \Lambda/3$, where $\Lambda$ is the cosmological constant, these modes are completely identical to the de Sitter modes \cite{Lopez-Ortega:2012xvr}. In fact, if both acceleration and the cosmological constant were present on the black hole, the end results would be exactly the same, except that the surface gravity would be given by
\begin{equation}
    k_A=-\sqrt{A^2+\Lambda/3}.
\end{equation}

\subsection{Accelerating Bertotti-Robinson modes}\label{Accelearting BR QNM}
%%%%%%%%%%%%%%%%%%%%%%%%%%%%%%%%%%%%%%%%%%%%%%%%%%%%%%%%%%%%%%%%%%%%%%
%%%%%%%%%%%%%%%%%%%%%%%%%%%%%%%%%%%%%%%%%%%%%%%%%%%%%%%%%%%%%%%%%%%%%%
\noindent
We now take back $M,Q\neq0$. For the regular accelerating Bertotti-Robinson regime, we consider $r_-\sim r_+\sim M<r_A$. This allows us to take the limit
\begin{equation}
    \frac{r_--r_+}{r_+}\ll 1.
\end{equation}
Given this, we can approximate the $f(r)$ function to 
\begin{equation}
    f(r)= \frac{(r-r_+)(r-r_-)(1-Ar)(1+Ar)}{r^2}\sim \frac{(r_A^2-r_+^2)}{r_A^2r_+^2}(r-r_+)(r-r_-).
    \label{f(r)BR}
\end{equation}
The surface gravity at the horizon $r_+$ becomes approximately
\begin{equation}
    k_+=\frac{(r_A^2-r_+^2)(r_+-r_-)}{2r_A^2r_+^2}.
    \label{k+BR}
\end{equation}
We now write the tortoise coordinate $x$ as a function of $r$, later invert it to get $r(x)$ and then plug it in \eqref{f(r)BR}.
This leads to 
\begin{equation}
    x \sim \frac{r_+^2}{1-A^2r_+^2}\int \frac{1}{(r-r_+)(r-r_-)}dr=\frac{1}{2k_+}\ln{\left(\frac{r-r_-}{r-r_+}\right)}.
\end{equation}
By inverting this relation we now obtain 
\begin{equation}
    r(x)=\frac{r_+e^{2k_+ x}-r_-}{e^{2k_+x}-1},
    \label{r(x)BR}
\end{equation}
and therefore 
\begin{equation}
    f(r)=\frac{(r_A^2-r_+^2)(r_+-r_-)^2}{r_A^2r_+^2} \frac{1}{4\sinh^2(k_-x)}=\left(\frac{r_A^2r_+^2}{r_A^2-r_+^2}\right)\frac{k_+^2}{\sinh^2(k_+x)}.
\end{equation}
Consequently, the resulting potential is of modified P\"oschl Teller kind, where $_s\tilde V_0$ gives
\begin{equation}
    _0\tilde V_0= k_+^2\frac{\lambda_0-1/3+A^2Q^2/3}{1-A^2M^2}, \quad _1\tilde V_0=k_+^2\frac{\lambda_1}{1-A^2M^2},\quad {}_{1/2}\tilde V_0=k_+^2\frac{\lambda^2_{1/2}}{1-A^2M^2}.
    \label{BRPotentials}
\end{equation}
%where we have renamed $\lambda_{1/2}$ as the square root of the $\lambda_{1/2}$ defined in section \ref{The Dirac field}.

The main problem here is that the resulting potential generically depends on $\lambda_s$. In Kofro\v{n}'s work \cite{Kofron:2015gli}, a solution for the separation constant of the general spin angular equation \eqref{general angular equation} is obtained, but only for vanishing azimuthal angular momentum number, $m=0$, and within the near extremal Bertotti-Robinson limit, $r_-\sim r_+<r_A$. Moreover, the differential equation considered in \cite{Kofron:2015gli} is not exactly of the same form as the one considered here, so some differences in the separation constant are to be expected when compared with our results below. We will now explain how this separation constant can be obtained using analytical methods similar to those employed for the radial potential.

Let us first recall that $r_-\sim r_+\sim M$. Furthermore, we will focus only on the scalar and electromagnetic fields, since the treatment of the fermionic field is more involved. In addition, the fermionic field only affects the real part of the quasinormal frequency, which will not be relevant for section \ref{Stability of the Cauchy horizon}.
We introduce the coordinate $u=\cos{\theta}$ and the variables $a=r_+/r_A$, $b=r_-/r_A$. This leads to 
\begin{equation}
    P(\theta)=ab(u-1/a)(u-1/b).
\end{equation}

In analogy to the treatment of the radial equation, we approximate $u\sim 1/a$. Thus the angular tortoise coordinate can be obtained
\begin{equation}
    z\sim \frac{a^2}{(1-a^2)}\int \frac{du}{P(u)}= \frac{a^2}{(1-a^2)(b-a)}\ln\left(\frac{u-1/a}{u-1/b}\right).\label{angulartortoiseBertottiRobinson}
\end{equation}
We define 
\begin{equation}
    k=\frac{(1-a^2)(b-a)}{2a^2}=\frac{(r_A^2-r_+^2)(r_--r_+)}{2r_+^2r_A}=-k_+r_A,
    \label{kei}
\end{equation}
where $k_+$ is defined in \eqref{k+BR}. Inverting \eqref{angulartortoiseBertottiRobinson} gives us
\begin{equation}
    u=\frac{1/a-e^{2kz}/b}{1-e^{2kz}}.
\end{equation}
Returning this to the $P(\theta)$ function we find that 
\begin{equation}
    P(\theta)=-\frac{b}{a} \frac{(1-a/b)^2}{4\sinh^2(kz)}.
\end{equation}
The resulting angular potentials becomes 
\begin{equation}
    _0V_\theta\sim\frac{k^2}{\sinh^2(kz)}\frac{a}{b(a^2-1)}\left(\lambda_0-\frac{1}{3}+\frac{A^2Q^2}{3}\right),\quad \,_1V_\theta\sim\frac{k^2}{\sinh^2(kz)}\frac{a}{b(a^2-1)}\lambda_1.
\end{equation}
Substituting in $a$ and $b$, we find that this gives the exact same potentials as described in \eqref{BRPotentials}, except by an overall minus sign and that $k_+$ is replaced by $k$ defined in equation \eqref{kei}. Notice that the radial and angular equations of motion \eqref{general radial equation} \eqref{general angular equation} can be identified by $\omega^2\to-m^2$ and $V_r\to -V_\theta$. We can now use the solutions of the P\"oschl Teller potential in the angular equation and therefore obtain $\lambda_s$. We get that
\begin{equation}
    im=i|k|\left(\sqrt{\frac{_s\tilde V_0}{k_+^2}+\frac{1}{4}}-\left(\ell+\frac{1}{2}\right)\right),
    \label{lambdasBR}
\end{equation}
where $\ell$ is the angular momentum number\footnote{The allowed values for $\ell$ is different depending on scalar and electromagnetic perturbations. While scalar fields allows $\ell=0$, electromagnetic fields do not. Since an $\ell=0$ mode has a resulting $_1V_r=0$, it therefore allows only the trivial vanishing field solution, $\phi_1=0$.} and $_s\tilde V_0$ is defined in \eqref{BRPotentials}. It naturally follows that 
\begin{equation}
    \sqrt{\frac{_s\tilde V_0}{k_+^2}+\frac{1}{4}}=\ell+\frac{1}{2}+\frac{m}{|k|}.
    \label{massage}
\end{equation}
Finally, the QNM frequencies and the GBF are readily calculable following the description in subsection \ref{Formalism}. They yield 
\begin{equation}
    \omega=-i(k_+\left(\ell+n+1 \right)+A|m|),
    \label{omegaBRnormal}
\end{equation}
\begin{equation}
\Gamma(\omega)=\frac{\sinh^2\left(\frac{\pi\omega}{k_+}\right)}{\sinh^2\left(\frac{\pi\omega}{k_+}\right)+\cos^2\left[\pi\left(\ell+\frac{1}{2}+\frac{A}{k_+}|m|\right)\right]}.
\label{gammaBRnormal}
\end{equation}
The absolute value of $m$ is necessary, because as we have seen in section \ref{Mode stability}, there should be no unstable modes with $\omega_I\geq0$. Equations \eqref{omegaBRnormal} and \eqref{gammaBRnormal} show that the fields with spin $s=0$ and $s=1$ have the same QNM frequency spectra and GBF\footnote{This statement holds for every $\ell\geq1$, since electromagnetic fields vanish for $\ell=0$, whereas scalar fields do not. The same convention will be used in the following sections without being stated explicitly.}. Furthermore, we now find an $m$ dependent term which vanishes identically in the $A\to0$ limit. As expected since this limit restores the spherical symmetry of the spacetime. Interestingly, the $m$ dependence of these modes is linear, which goes in contrast with the $m^2$ dependence found for the photon surface family in the small acceleration limit \cite{Moura:2026ken}.

If our goal is to determine $\lambda_s$, we only need to rearrange \eqref{massage} to obtain
\begin{equation}
\begin{aligned}
    &\lambda_0 \sim \left(\ell(\ell+1)+2\left(\ell+\frac{1}{2}\right)\frac{Am}{k_+}+\frac{A^2m^2}{k_+^2}+\frac{1}{3}\right)(1-A^2M^2),  \\   &\lambda_1\sim \left(\ell(\ell+1)+2\left(\ell+\frac{1}{2}\right)\frac{Am}{k_+}+\frac{A^2m^2}{k_+^2}\right)(1-A^2M^2).
\end{aligned}
    \label{sepconstnatBertottiRobinson}
\end{equation}
\subsection{Accelerating Nariai modes}\label{accelerating Nariai QNM}
%%%%%%%%%%%%%%%%%%%%%%%%%%%%%%%%%%%%%%%%%%%%%%%%%%%%%%%%%%%%%%%%%%%%%%
%%%%%%%%%%%%%%%%%%%%%%%%%%%%%%%%%%%%%%%%%%%%%%%%%%%%%%%%%%%%%%%%%%%%%%
\noindent
For the regular accelerating Nariai regime we consider $r_-<r_+\sim r_A$ and therefore take the approximation
\begin{equation}
0<\frac{r_+-r_A}{r_A}\ll 1.
\label{conditionNariai}
\end{equation}
We have imposed that the quantity in \eqref{conditionNariai} is positive, although it could equally well be negative. This choice simply fixes $r_+$ to be the black hole event horizon; equivalently, one could instead choose $r_A$ to play this role if one would rather consider this limit coming from the super accelerating regime.

The procedure is similar to what followed in the previous section \ref{Accelearting BR QNM}. The function $f(r)$ can now be approximated by
\begin{equation}
\begin{aligned}
        f(r)= \frac{(r-r_+)(r-r_-)(1-Ar)(1+Ar)}{r^2}\sim& \frac{(r_A+r_+)(r_+-r_-)}{r_A^2r_+^2}(r-r_+)(1/A-r)\sim  \\ \sim &\frac{2(r_+-r_-)}{r_+^3}(r-r_+)(r-r_A),
\end{aligned}
\end{equation}
and its surface gravity is
\begin{equation}
    k_+=\frac{(r_A+r_+)(r_+-r_-)}{2r_A^2r_+^2}\sim\frac{(r_A-r_+)(r_+-r_-)}{r_+^3}.
    \label{k+Nariai}
\end{equation}
Consequently, we obtain $r$ as a function of the tortoise $x$ to be 
\begin{equation}
    r(x)=\frac{r_Ae^{2k_+x}+r_+}{1+e^{2k_+x}}
\end{equation}
which resembles equation \eqref{r(x)BR}. By plugging this back into $f(r)$ we get
\begin{equation}
    f(x)=\frac{(r_A-r_+)^2}{4\cosh^2{(k_+x)}}\frac{2(r_+-r_-)}{r_+^3} \sim \frac{r_+^3}{2(r_+-r_-)}\frac{k_+^2}{\cosh^2{(k_+x)}}.
\end{equation}
So we can easily obtain the P\"{o}schl Teller potential with constant $_sV_0$ to be given by
\begin{equation}
    _0V_0\sim k_+^2\left(\lambda_0\frac{r_+}{2(r_+-r_-)}+\frac{1}{3}\right),\quad _1V_0\sim k_+^2\lambda_1\frac{r_+}{2(r_+-r_-)},\quad _{1/2}V_0\sim k_+^2\lambda^2_{1/2}\frac{r_+}{2(r_+-r_-)}.
    \label{potentialsNariai}
\end{equation}
%where we have renamed $\lambda_{1/2}$ as the square root of the $\lambda_{1/2}$ defined in section \ref{The Dirac field}.

Again, we need to obtain an expression for the separation constant $\lambda_s$. We will operate similarly as we have done in \ref{Accelearting BR QNM}, where we perform an identical analysis for the radial potential.

We once more define $a=r_+/r_A$, $b=r_-/r_A$ and take $u=\cos\theta$. In this spacetime regime, the angular tortoise is 
\begin{equation}
    z=-\int\frac{du}{(1-u)(1+u)(au-1)(bu-1)}\sim \frac{1}{2a(1-b)}\int \frac{du}{(u-1/a)(1-u)}.
\end{equation}
This yields
\begin{equation}
    z=\frac{1}{2k}\ln\left(\frac{u-1/a}{1-u}\right),
    \label{angulartortoiseNariai}
\end{equation}
where we have defined
\begin{equation}
    k=(a-1)(1-b)=-k_+r_A,
\end{equation}
with $k_+$ defined in \eqref{k+Nariai}. Inverting \eqref{angulartortoiseNariai} gives
\begin{equation}
    u=\frac{1/a+e^{2kz}}{1+e^{2kz}}.
\end{equation}
We can now compute
\begin{equation}
    P(u)(1-u^2)=\frac{(b-1)(1-a)^2}{2\cosh^2(kz)},
\end{equation}
and therefore obtain the angular potential function
\begin{equation}
    _0V_\theta=\frac{k^2}{2a(b-1)\cosh^2(kz)}\left( \lambda_0+\frac{2}{3}\frac{(r_+-r_-)}{r_+}\right),\quad _1V_\theta=\frac{k^2}{2a(b-1)\cosh^2(kz)} \lambda_1.
\end{equation}
We therefore observe that this gives exactly the same $_sV_0$ described by \eqref{potentialsNariai}, but with an overall minus sign and different $k$ constant. Consequently, thorough the use of the same argument as in the previous section \ref{Accelearting BR QNM}, we find that
\begin{equation}
    i|m|=i|k|\left( \sqrt{\frac{_sV_0}{k_+^2}-\frac{1}{4}}-\left(\ell+\frac{1}{2}\right)\right),
\end{equation}
where $_sV_0$ is defined in \eqref{potentialsNariai}. This allows us to conclude that
\begin{equation}
    \sqrt{\frac{_sV_0}{k_+^2}-\frac{1}{4}}=\ell+\frac{1}{2}+\frac{m}{|k|}.
    \label{massage2}
\end{equation}
This gives the quasinormal mode frequencies
\begin{equation}
    \omega =\pm mA+k_+\left(\pm\left(\ell+\frac{1}{2}\right)-i\left( n+\frac{1}{2}\right) \right)
\end{equation}
and the greybody factor
\begin{equation}
\Gamma(\omega)=\frac{\sinh^2\left(\frac{\pi\omega}{k_+}\right)}{\sinh^2\left(\frac{\pi\omega}{k_+}\right)+\cosh^2\left[\pi\left(\ell+\frac{1}{2}+\frac{A}{k_+}m\right)\right]}.
\end{equation}
Once more, we see that the lack of spherical symmetry of the black hole has an observable effect coming from the $m$ dependence of the quasinormal modes and greybody factors.

Moreover, if one whishes to obtain the separation constant, we simply have to massage equation \eqref{massage2} to obtain
\begin{equation}
\begin{aligned}
    &\lambda_0\sim\frac{2(r_+-r_-)}{r_+}\left(\ell(\ell+1)+2\left(\ell+\frac{1}{2}\right)\frac{Am}{k_+}+\frac{A^2m^2}{k_+^2}-\frac{1}{3}\right),
    \\
    &\lambda_1\sim\frac{2(r_+-r_-)}{r_+}\left(\ell(\ell+1)+2\left(\ell+\frac{1}{2}\right)\frac{Am}{k_+}+\frac{A^2m^2}{k_+^2}\right).
\end{aligned}
\end{equation}

\subsection{Super accelerating Bertotti-Robinson modes}\label{Super accelerating Bertotti-Robinson modes}%%%%%%%%%%%%%%%%%%%%%%%%%%%%%%%%%%%%%%%%%%%%%%%%%%%%%%%%%%%%%%%%%%%%%%
%%%%%%%%%%%%%%%%%%%%%%%%%%%%%%%%%%%%%%%%%%%%%%%%%%%%%%%%%%%%%%%%%%%%%%
\noindent
In the super accelerating Bertotti-Robinson regime, we consider $r_A\sim r_-<r_+$, such that
\begin{equation}
    0<\frac{r_A-r_-}{r_-}\ll1.
    \label{conditionsuperaccbertotti}
\end{equation}
We have taken the quantity in \eqref{conditionsuperaccbertotti} to be positive, although the same analysis could equally be carried out when it is negative. This choice simply fixes $r_-$ as the event horizon.

The horizon function $f(r)$ is then given by
\begin{equation}
    f(r)\sim\frac{(r_--r_+)(r_-+r_A)}{r_-^2r_A^2}(r-r_-)(r_A-r)\sim \frac{2(r_--r_+)}{r_-^3}(r-r_-)(r_A-r),
\end{equation}
and its surface gravity
\begin{equation}
    k_-\sim\frac{(r_--r_+)(r_-+r_A)(r_A-r_-)}{2r_-^2r_A^2}\sim \frac{(r_--r_+)(r_A-r_-)}{r_-^3}.
\end{equation}
At this point, we notice that these expressions are identical to those obtained in the accelerating Nariai case discussed in Section \ref{accelerating Nariai QNM}, with $r_+$ and $r_-$ interchanged.
Consequently, the resulting potential is of P\"{o}schl Teller kind with
\begin{equation}
    _0V_0=\frac{r_-}{2(r_--r_+)}\lambda_0+\frac{1}{3},\quad _1V_0=\frac{r_-}{2(r_--r_+)}\lambda_1,\quad _{1/2}V_0=\frac{r_-}{2(r_--r_+)}\lambda^2_{1/2},
\end{equation}
which is generically negative.

To obtain the separation constant, we can follow the analogous procedure for the angular equation which gives
\begin{equation}
\begin{aligned}
    &\lambda_0\sim\frac{2(r_+-r_-)}{r_-}\left( \ell(\ell+1) +2\left( \ell+\frac{1}{2}\right)\frac{Am}{k_-}+\frac{A^2m^2}{k_-^2}-\frac{1}{3}\right), \\
    &\lambda_1\sim\frac{2(r_+-r_-)}{r_-}\left( \ell(\ell+1) +2\left( \ell+\frac{1}{2}\right)\frac{Am}{k_-}+\frac{A^2m^2}{k_-^2}\right).
\end{aligned}
\end{equation}
The QNM frequencies are thus 
\begin{equation}
    \omega=-i(k_-\left(\ell+n+1 \right)+|m|A),
\end{equation}
and the GBF is
\begin{equation}
\Gamma(\omega)=\frac{\sinh^2\left(\frac{\pi\omega}{k_+}\right)}{\sinh^2\left(\frac{\pi\omega}{k_+}\right)+\cos^2\left[\pi\left(\ell+\frac{1}{2}+\frac{|m|}{k_-}A\right)\right]}.
\end{equation}

\subsection{Super accelerating Nariai modes}\label{Super accelerating Nariai modes}
%%%%%%%%%%%%%%%%%%%%%%%%%%%%%%%%%%%%%%%%%%%%%%%%%%%%%%%%%%%%%%%%%%%%%%
%%%%%%%%%%%%%%%%%%%%%%%%%%%%%%%%%%%%%%%%%%%%%%%%%%%%%%%%%%%%%%%%%%%%%%
\noindent
In the super accelerating Nariai spacetime, we have $r_A<r_-\sim r_+$. We notice that this configuration is identical to the regular accelerating Bertotti-Robinson one described in section \ref{Accelearting BR QNM}, except that $r_+$ and $r_-$ are to be swapped. 
Therefore, the $f(r)$ function becomes
\begin{equation}
    f(r)\sim \frac{2(r_+-r_-)}{r_-^3}\left(r-r_- \right)(r_A-r)
\end{equation}
and the horizon surface gravity 
\begin{equation}
    k_-\sim \frac{(r_+-r_-)(r_A-r_-)}{r_-^3}.
\end{equation}
Consequently, the $_sV_0$ expressions found in equation \eqref{BRPotentials} are identical here, but now negative. Additionally, the separation constant becomes similar to what is given by equation \eqref{sepconstnatBertottiRobinson}, where we obtain
\begin{equation}
\begin{aligned}
    &\lambda_0 \sim \left(\ell(\ell+1)+2\left(\ell+\frac{1}{2}\right)\frac{Am}{k_-}+\frac{A^2m^2}{k_-^2}+\frac{1}{3}\right)(A^2M^2-1),
    \\ &\lambda_1\sim \left(\ell(\ell+1)+2\left(\ell+\frac{1}{2}\right)\frac{Am}{k_-}+\frac{A^2m^2}{k_-^2}\right)(A^2M^2-1).
\end{aligned}
\end{equation}
The quasinormal frequencies, together with the greybody factor are then given by
\begin{equation}
    \omega =\pm mA+k_-\left(\pm\left(\ell+\frac{1}{2}\right)-i\left( n+\frac{1}{2}\right) \right),
\end{equation}
and
\begin{equation}
\Gamma(\omega)=\frac{\sinh^2\left(\frac{\pi\omega}{k_+}\right)}{\sinh^2\left(\frac{\pi\omega}{k_+}\right)+\cosh^2\left[\pi\left(\ell+\frac{1}{2}+\frac{m}{k_-}A\right)\right]}.
\end{equation}

\section{Stability of the Cauchy horizon and strong cosmic censorship}\label{Stability of the Cauchy horizon}
%%%%%%%%%%%%%%%%%%%%%%%%%%%%%%%%%%%%%%%%%%%%%%%%%%%%%%%%%%%%%%%%%%%%%%
%%%%%%%%%%%%%%%%%%%%%%%%%%%%%%%%%%%%%%%%%%%%%%%%%%%%%%%%%%%%%%%%%%%%%%
\noindent
Given that we have already computed the quasinormal mode frequencies associated to each quasinormal mode family of accelerating and super accelerating black holes, we are now ready to tackle the problem of the stability of the Cauchy horizon and the strong cosmic censorship conjecture.

We have established in the introduction \ref{Introduction}, the modern version of strong cosmic censorship. Following up, Christodoulou also provided a statement which allows to evaluate whether such conjecture holds \cite{Christodoulou:2008nj}. It includes looking at the quasinormal modes of the spacetime in question, as they provide information on how perturbations grow. We call his criterion, Christodoulou's $\beta$ criterion and it goes as follows. First define the spectral gap $\alpha$ as the infimum of minus the imaginary part of the quasinormal mode frequencies
\begin{equation}
    \alpha=\inf [-\omega_I];
\end{equation}
then let $k_C$ be the surface gravity of the Cauchy horizon. Finally, obtain $\beta$ by
\begin{equation}
    \beta=\frac{\alpha}{|k_C|}.
    \label{beta}
\end{equation}
Then Christodoulou's $\beta$ criterion claims \cite{Christodoulou:2008nj}: if 
\begin{equation}
    \beta<\frac{1}{2},
\end{equation}
then the Cauchy horizon is unstable and strong cosmic censorship is preserved\footnote{The applicability of this criterion relies on some non trivial assumptions. In \cite{Destounis:2020yav}, it is argued that these assumptions are well justified for the C-metric. Essentially, it is due to the fact that, analogously to Reissner N\"{o}rdstrom de Sitter black holes, the C-metric also exhibits a cosmological horizon. As a consequence, late time perturbations decay exponentially rather than by a power law as described by Price's law \cite{Price:1971fb}. This kind of decay is key, since it means that late time tails may decay fast enough to counterbalance the exponential blueshift at the Cauchy horizon, which is the source of the instability.}. If one finds otherwise that $\beta\geq1/2$, it means that SCC may be violated. However, it is not conclusive since back reaction or quantum effects can play an important role on this problem. Additionally, non smooth (rough) initial data can also contribute on the preservation of SCC \cite{Dias:2018etb}. Nevertheless, we will not discuss these issues further in this work.

The path ahead is now clear. Our goal will be to search through all the QNM families and through all the black hole parameters, therefore finding if there is any possibility that $\beta\geq1/2$. Additionally, we stress that the interest of this work, related with other works on the stability of the Cauchy horizon of the C-metric \cite{Destounis:2020yav}, is that we are now considering the wider class of solutions which contain the super accelerating regime. In fact, for the first super accelerating regime, the horizon labeled by acceleration $r_A=1/A$ becomes the event horizon of the black hole, while in the second super accelerating regime, it becomes the Cauchy horizon. Thus we ask, is there any special property that this swapping of horizons introduces in the stability of the Cauchy horizon problem?

\subsection{Photon surface family}
%%%%%%%%%%%%%%%%%%%%%%%%%%%%%%%%%%%%%%%%%%%%%%%%%%%%%%%%%%%%%%%%%%%%%%
%%%%%%%%%%%%%%%%%%%%%%%%%%%%%%%%%%%%%%%%%%%%%%%%%%%%%%%%%%%%%%%%%%%%%%
\noindent
Let us begin by considering the photon surface family. We emphasize once more that the photon surface modes are generically valid over a wide range of black hole parameters $M,Q,A$. This means that, in the purely accelerating and near extremal limits, the photon surface modes continuously deform into the corresponding modes of these other families. On the other hand, as one approaches these limits, the QNM frequencies are more accurately described by the expressions obtained directly for the respective limiting families.

Further, it is worth analyzing how good of an approximation the photon surface modes give for the $\beta$ computation in \eqref{beta}. From the QNM frequencies, $\beta$ relies only on their imaginary part. We have prescribed that photon surface modes are valid in the limit where the separation constant $\lambda\to \infty$; moreover, we found that by use of WKB methods \eqref{wkb2}, $\omega_I$ does not depend on $\lambda$ to first order. Further, additional correction are to be of order of negative powers of $\lambda$ \cite{Cardoso:2008bp}, which are negligible in the large $\lambda$ limit. Therefore, we find that generically the WKB expression \eqref{wkb2} has two big advantages. One, we do not need to calculate the actual value of $\lambda$; two it provides a good approximation in the eikonal limit. This last one has been confirmed by comparing the expression with numerical calculations for the regular accelerating regime in \cite{Moura:2026ken}.

%%%%%%%%%%%%%%%%%%%%%%%%%%%%%%%%%
Concluding, we just need to focus on the formula obtained for the imaginary part of the QNM frequency \eqref{wkb2}. The spectral gap $\alpha$ is easily obtained by setting the overtone number $n=0$ and we can now scan the entire parameter space to find possible violations of Christodoulou's criterion. We can now scan the entire parameter space to search for possible violations of Christodoulou's criterion. To this end, we plot a two dimensional surface embedded in three dimensional Euclidean space. The quantity $\beta$ is plotted along the $z$ axis, while the ratio of the event horizon radius to the cosmological horizon radius is plotted along the $x$ axis, and the ratio of the Cauchy horizon radius to the event horizon radius along the $y$ axis. For reference, we also plot the plane $z=1/2$.

\begin{figure}[htp]
\centering
\includegraphics[width=.5\textwidth]{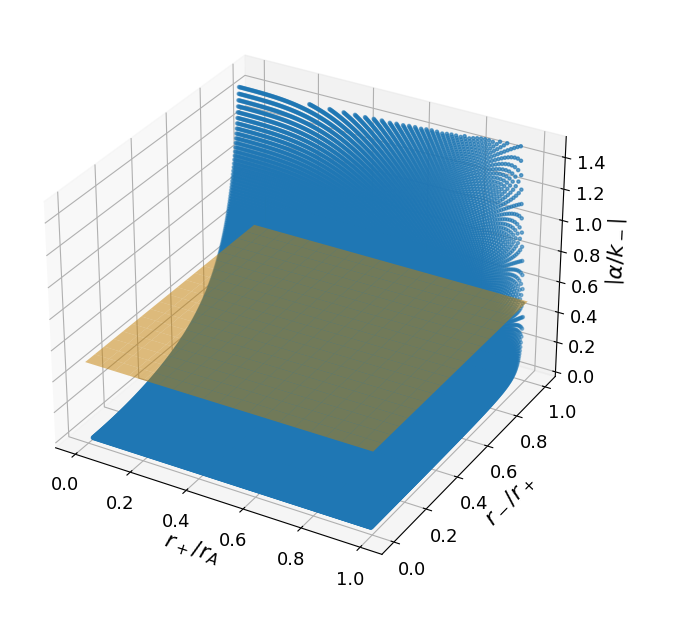}\hfill
\includegraphics[width=.5\textwidth]{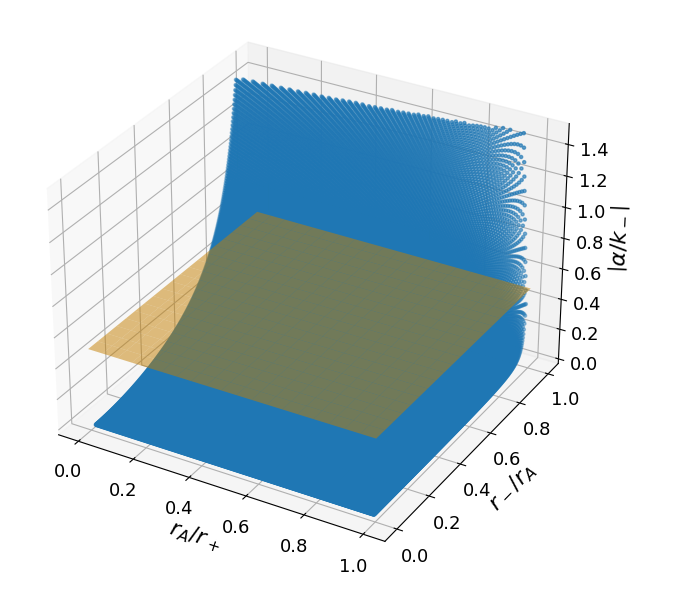}\hfill
\includegraphics[width=.5\textwidth]{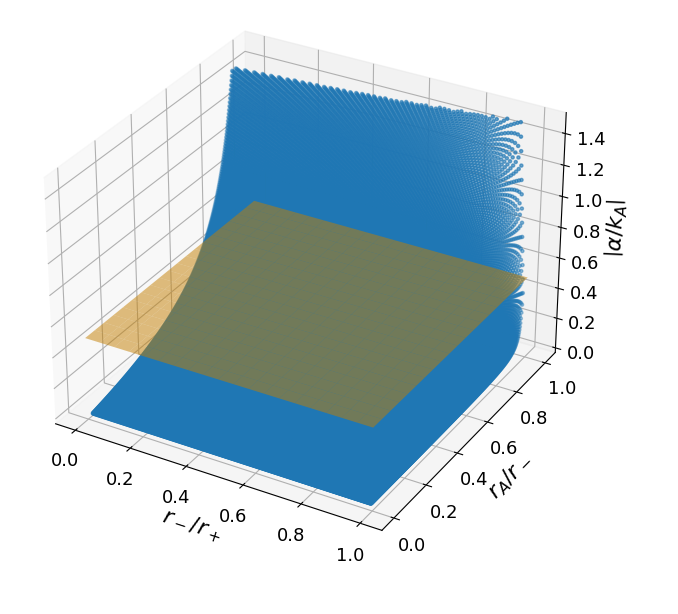}
\caption{Parameter $\beta$ \eqref{beta} computed for all the parameter space of the accelerating and super accelerating C-metric. On the $x$ axis we have the radius of the event horizon over the radius of the cosmological horizon and on the $y$ axis we have the Cauchy horizon over the event horizon. The three plots represent the three spacetime regimes: from left to right, first is the regular accelerating regime ($r_-<r_+<r_A$), second is the first super accelerating regime ($r_-<r_A<r_+$) and third (below) is the second super accelerating regime ($r_A<r_-<r_+$). The plane $\beta=1/2$ is plotted for comparison.}
\label{Photon Sphere WKB}
\end{figure}

Since the labels assigned to the horizons change between the different accelerating regimes, we perform this analysis separately for each regime, resulting in three plots. In this way, we can scan the full range of black hole parameters, with the $x$ and $y$ coordinates varying from $0$ to $1$. The resulting plots are shown in figure \ref{Photon Sphere WKB}.

By construction, the plots in figure \ref{Photon Sphere WKB} represent the Nariai family (either super or regular accelerating) when $x\sim 1$, while the Bertotti-Robinson family (either super or regular accelerating) is found when $y\sim1$. From figure \ref{Photon Sphere WKB}, we see that, throughout most of the parameter space, the photon surface modes preserve Christodoulou's version of SCC. However, there is a small region where $\beta\geq1/2$. This occurs when the black hole approaches the Bertotti-Robinson limit, in both the regular and super accelerating regimes. We are therefore led to examine these mode families in more detail. In this regime, we expect the results obtained in subsections \ref{Accelearting BR QNM} and \ref{Super accelerating Bertotti-Robinson modes} to provide more accurate QNM frequencies than those obtained from the WKB approximation considered here.

Before proceeding, note that figure \ref{Photon Sphere WKB} also shows that $\beta$ does not tell us much about what happens for the Nariai family, nor does it gives much information regarding purely accelerating modes. So without further ado, let us analyze them as a preliminary.

\subsection{Purely accelerating family}
%%%%%%%%%%%%%%%%%%%%%%%%%%%%%%%%%%%%%%%%%%%%%%%%%%%%%%%%%%%%%%%%%%%%%%
%%%%%%%%%%%%%%%%%%%%%%%%%%%%%%%%%%%%%%%%%%%%%%%%%%%%%%%%%%%%%%%%%%%%%%
\noindent
Starting with the purely accelerating family, we have noted in subsection \ref{Purely accelerating modes} that these modes are purely damped and completely determined by the surface gravity of the Rindler acceleration horizon, defined as $k_A^R=A$. To obtain these modes, we considered $M=Q=0$. Nevertheless, this family continues to exist for general $M,Q\neq0$, although the modes become increasingly deformed as $M$ and $Q$ increase. Consequently, they must cease to exist in any near extremal regime.

Using the expression obtained in subsection \ref{Purely accelerating modes}, we can analyze the behavior of $\beta$. We find
\begin{equation}
\beta\propto \frac{k_A^R}{|k_-|}.
\end{equation}
It follows that violations of SCC can occur whenever $|k_-|$ is sufficiently small. In the Bertotti-Robinson limit, the surface gravity satisfies $|k_-|\to0$; consequently $\beta\to\infty$. So we find that, in fact,
this mode family can violate Christodoulou’s criterion of SCC as long as it survives the deformations mentioned above.

\subsection{Nariai family}\label{Nariai family}
%%%%%%%%%%%%%%%%%%%%%%%%%%%%%%%%%%%%%%%%%%%%%%%%%%%%%%%%%%%%%%%%%%%%%%
%%%%%%%%%%%%%%%%%%%%%%%%%%%%%%%%%%%%%%%%%%%%%%%%%%%%%%%%%%%%%%%%%%%%%%
\noindent
Let us now consider the Nariai family. The following analysis applies generically to both the regular and super accelerating regimes. As found in sections \ref{accelerating Nariai QNM} and \ref{Super accelerating Nariai modes}, the imaginary part of the QNM frequencies, $\omega_I$, is proportional to the surface gravity of the black hole event horizon, $k_H$.

In the near extremal regime, as the radius of the event horizon ($r_H$) approaches that of the cosmological horizon ($r_c$), the surface gravity of the event horizon goes to zero ($k_H\to0$). On the other hand, the surface gravity of the Cauchy horizon ($k_C$) remains finite and non zero. Consequently,
\begin{equation}
\beta \propto \frac{k_H}{|k_C|}\ll 1.
\end{equation}
We therefore conclude that the Nariai family cannot violate Christodoulou's version of strong cosmic censorship.

\subsection{Bertotti-Robinson family}
%%%%%%%%%%%%%%%%%%%%%%%%%%%%%%%%%%%%%%%%%%%%%%%%%%%%%%%%%%%%%%%%%%%%%%
%%%%%%%%%%%%%%%%%%%%%%%%%%%%%%%%%%%%%%%%%%%%%%%%%%%%%%%%%%%%%%%%%%%%%%
\noindent 
We shall now deal with both the super and regular accelerating Bertotti-Robinson spacetimes whose QNM frequencies were computed in section \ref{Super accelerating Bertotti-Robinson modes} and \ref{Accelearting BR QNM}, respectively. There, we can easily identify that the dominant mode for scalar and electromagnetic perturbations is obtained when the azimuthal angular number is vanishing, $m=0$. In the fermionic case, $\omega_I$ does not depend on $m$ in any way. So without loss of generality, let us take $m=0$.

Identically to the Nariai families, we find that the spectral gap of these modes is proportional to the event horizon surface gravity $k_H$. Moreover, we have $k_H\to 0$ as $r_H\to r_C$, where $r_H$ denotes the radius of the event horizon and $r_C$ the radius of the Cauchy horizon. However, the same also happens to the surface gravity of the Cauchy horizon: $|k_C|\to0$ as $r_H\to r_C$. Consequently, it is possible to establish the following result:
\begin{equation}
    \frac{k_H}{|k_C|}\lesssim 1,
    \label{k_H/|k_C| BR}
\end{equation}
where the given ration approaches $1$ as the black hole approaches extermality. The proof to this relation should be performed for each accelerating regime separately. Consider first the regular accelerating regime, where $r_H=r_+$ and $r_C=r_-$. We have that
\begin{equation}
    \frac{k_+}{|k_-|}=\left(\frac{r_-}{r_+}\right)^2\frac{r_A^2-r_+^2}{r_A^2-r_-^2}.
\end{equation}
Since in this regime $r_-<r_+<r_A$, we are led to conclude that $k_+/|k_-|\le 1$. Equality is established in the exact regular accelerating extremal Bertotti-Robinson limit.
For the first super accelerating regime, $r_H=r_A$ and $r_C=r_-$, therefore
\begin{equation}
    \frac{k_A}{|k_-|}=\frac{r_+-r_A}{r_+-r_-}\frac{2r_-^2}{r_A(r_A+r_-)}.
    \label{kA/|k-|}
\end{equation}
Since $r_-<r_A<r_+$, we conclude that $k_A/|k_-|\leq1$ and the equality is attained in the actual super accelerating Bertotti-Robinson extremal limit. On the second super accelerating regime, $r_H=r_-$ and $r_C=r_A$, we have to consider the ratio
\begin{equation}
    \frac{k_-}{|k_A|}=\frac{r_+-r_-}{r_+-r_+}\frac{r_A(r_A+r_-)}{2r_-^2}
\end{equation}
and analogously to \eqref{kA/|k-|}, we obtain that $k_-/|k_A|\leq1$, with equality obtained in the actual super accelerating Bertotti-Robinson extremal limit.

The property found in equation \eqref{k_H/|k_C| BR} is crucial to establish upper bounds on $\beta$. Let us start by computing $\beta$ for the massless Dirac fields. In the previous section \ref{Quasinormal mode families}, we saw that for near extremal spacetimes, fermions have QNM frequencies whose imaginary part is
\begin{equation}
    \omega_I=-k_H\left(n+\frac{1}{2}\right),\quad n=0,1,\dots.
    \label{diracwi}
\end{equation}
The spectral gap is obtained by the negative of \eqref{diracwi} and taking $n=0$. Therefore we find that 
\begin{equation}
    \beta=\frac{k_H}{2|k_C|}.
\end{equation}
Since $k_H/|k_C|\lesssim 1$, we conclude that
\begin{equation}
    \beta \lesssim \frac{1}{2}.
\end{equation}
Consequently, massless Dirac fields do not violate strong cosmic censorship.

We now should find out what happens to the massless scalar and electromagnetic fields. Given that we have already taken $m=0$, to find the spectral gap we should search for $\ell$ and $n$ that establish the dominant mode. The QNM frequency is described by its imaginary part
%We can easily check from the expressions found in chapter \ref{ch:SL}, that zero azimuthal angular momentum number ($m=0$) modes are the dominate ones. From now on, let us take $m=0$ then. We get
\begin{equation}
    \omega_I^{(m=0)}=-k_H\left(n+\ell +1\right),\quad n=0,1,2\dots.
    \label{omegaIscalarelectro}
\end{equation}
Consequently, the spectral gap is obtained by taking the negative of \eqref{omegaIscalarelectro} and choosing the smallest allowed values of $n$ and $\ell$. We therefore take $n=0$. The smallest allowed value of $\ell$ depends on the type of perturbation: for the scalar field, we can take $\ell=0$, whereas for electromagnetic perturbations we must take $\ell=1$. This is because the potential ${}_1V_r$ vanishes for $\ell=0$ and therefore the wave equation only admits the trivial $\phi_1=0$ solution.

In the following, we can treat both cases simultaneously by taking $\ell=s$, where $s$ denotes the spin of the field.
This yields
\begin{equation}
    \beta=\frac{k_H}{|k_C|}(s+1).
\end{equation}
Let us first see how large $\beta$ can become. We use $k_H/|k_C|\lesssim 1$ to find that 
\begin{equation}
    \beta\lesssim s+1.
\end{equation}
This means that while electromagnetic perturbations cannot obtain $\beta$ greater than $2$, scalar perturbations do not get $\beta$ greater than $1$.

On the other hand, we still want to find violations of Christodoulou's criterion. This makes us look for constraints on the black hole parameters which satisfy $\beta\geq1/2$. Indeed, this condition imposes the following constraint
\begin{equation}
    k_H\geq \frac{|k_C|}{2(s+1)}.
\end{equation}
We observe that this constraint is easier to satisfy for scalar fields than for electromagnetic fields, for which the allowed parameter space is more restricted. Nevertheless, when sufficiently close to the extremal limit, where $k_H\sim |k_C|$, the inequality is satisfied for both types of perturbations. We therefore conclude that possible violations of Christodoulou's criterion for strong cosmic censorship occur in both the regular and super accelerating Bertotti-Robinson regimes, for both massless scalar and electromagnetic fields. Thus, the super accelerating regime does not alter the known result for strong cosmic censorship in the regular accelerating regime.

\section{Conclusion}\label{Conclusion}
%%%%%%%%%%%%%%%%%%%%%%%%%%%%%%%%%%%%%%%%%%%%%%%%%%%%%%%%%%%%%%%%%%%%%%
%%%%%%%%%%%%%%%%%%%%%%%%%%%%%%%%%%%%%%%%%%%%%%%%%%%%%%%%%%%%%%%%%%%%%%
\noindent
In this work, we have studied how field perturbations behave on accelerating and super accelerating black hole spacetimes, governed by the C-metric. We were lead to consider only conformally invariant fields as the conformal factor present in the metric complicates the separability of the equations of motion. Therefore, we were restricted to the massless scalar field, the massless Dirac field and the electromagnetic (Maxwell) field. Additionally, the massless spin $3/2$ (Rarita Schwinger) fermion field could have been considered, but we have decided to restrict the analysis to the simplest fermionic field here. We derived a master Sch\"{o}dinger-like equation for the radial sector of the field as well as for the angular sector. In fact, the lack of spherical symmetry on the spacetime provided that the separation of variables introduced a separation constant which is generically not known. In addition, it is worth mentioning that we have found the polar and axial modes of electromagnetic fields to satisfy the same radial and angular master equations and therefore, they become isospectral when subject to the same boundary conditions, similar to what happens in non accelerating spacetimes. Moreover, we found that massless Dirac fermions on the black hole background hide a supersymmetry algebra, that allows one to relate their different degrees of freedom. Consequently, their equations of motion come in super partners, and accordingly non zero modes share the same spectrum, leading to the phenomenon of isospectrality.

Having derived the master equations for the fields on the C-metric background, we proceeded to analyze their linear stability in the regular and super accelerating regimes. This entailed in inquiring whether squared integrable fields would necessary have a real oscillatory frequency $\omega$. As a consequence, we had to analyze whether the governing potential is positive definite and therefore, bounds on the separation constant had to be imposed. Later, by analyzing the angular master equation, these bounds were found to be true in both the accelerating and super accelerating regimes, with one exception: the scalar field on a given region parameter of the super accelerating regime. Neglecting this aspect, we find that not only the regular accelerating regime is linearly stable, so is the super accelerating regime too.

To prepare for the problem of quasinormal modes, it is important to note whether these perturbations are modally stable or not. We were thus led to review a known proof regarding mode stability, where we concluded that introducing the super accelerating regime does not affect this proof in any way. Hence we established that all the QNM frequencies cannot have positive imaginary frequency, accordingly describing stable damped oscillations.

Later, we studied the scattering problem of quasinormal mode and greybody factors of these fields on the C-metric. We found that previous classifications of the QNM families of the spacetime can be generalized to the super accelerating regimes. Namely, while on the regular regime we had: purely acceleration modes, photon surface modes and two near extremal modes (accelerating Nariai and accelerating Bertotti-Robinson); we found that there are additionally two near extremal families, concerning the new near extremal limits of the super accelerating regimes (super accelerating Nariai and super accelerating Bertotti-Robinson) and that the photon surface modes survive in the super accelerating regimes as well. This provides a total of six mode families\footnote{Counting the photon surface family only once.} for the total parameter space region of the C-metric. 

For each of these families, we find analytical methods for obtaining the QNM spectra and greybody factors. Therefore, we managed to obtain such expressions for the two quantities of interest. A difficulty encountered was that the final results were dependent on the separation constant, which is generically not known. We applied similar methods as the ones performed for the radial potential, but to the angular potential to get an expression for the separation constant valid in the near extremal limits of the spacetime. Thus, we were able to obtain the QNM frequencies and the GBF written solely in terms of the surface gravity of the event horizon of the black hole configuration in question, the black hole's acceleration $A$, as well as the quantum numbers $m$, $\ell$ and $n$. The $m$ dependence is a curious property that reflects the lack of spherical symmetry in the spacetime, but it naturally goes to zero in the limit of $A\to0$, where spherical symmetry is recovered.

Moreover, we should mention that while the super accelerating regime modifies the boundary conditions of the angular equation, due to the new axis of symmetry it presents \cite{Silva:2026tev}, we find that it does not change the nature of the underlying problem. Meaning, the angular equation still defines an hermitian eigenvalue problem and thus the separation constant remains real valued. Furthermore, we note that there are other possible ways to obtain these QNM frequencies or even the separation constant. One of such is by use of connection coefficients of Heun functions where recent advances have been obtained \cite{Bonelli:2022ten} and, for instance, has precisely been used in \cite{Capuano:2026tjy}. Another possibility is to consider the highly damped limit where techniques from complex analysis, such as Stokes phenomena and monodromy methods, can be employed. The corresponding calculation for Schwarzschild de Sitter black holes in \cite{Cardoso:2004up} provides a potentially useful starting point for extending these methods to the C-metric, as the horizon structure of the two spacetimes share similarities. This employment of this last method is the work of an upcoming article.

Following up with the calculation of the quasinormal modes frequencies, it was possible to analyze the stability of the Cauchy horizon, which goes hand in hand with the strong cosmic censorship conjecture. This was done using Christodoulou's $\beta$ criterion. We searched through all the quasinormal mode families and found possible violations of SCC present when the black hole is close to either the regular or super accelerating Bertotti-Robinson limits. These violations, however, extend only to scalar and electromagnetic perturbations, given that we have discovered that fermions do not violate Christodoulou's $\beta$ criterion. In this view, these results are identical to the Reissner N\"{o}rdstrom de Sitter black hole, in the sense that violations to the $\beta$ criterion are also found in the Bertotti-Robinson limit. Additionally, we also find that the super accelerating regime, which is characterized by having the horizon labeled by acceleration $r_A=1/A$ playing the role of the event horizon of the black hole or even the Cauchy horizon, does not change the resulting story, indicating that this horizon works perfectly well as any other. Nonetheless, these findings suggest that predictability may be destroyed in general relativity which is a major conceptional question regarding the foundation of the theory. On the other hand, these results do not provide the final answer to the problem, as back reaction and quantum effects may play important effects on the corresponding problem, as well as considering non smooth initial data.

We leave several possible avenues for future investigations of field perturbations on the C-metric. In particular, several aspects of the generalized Teukolsky equation on accelerating and rotating black hole backgrounds remain unexplored. In the Kerr spacetime, the Teukolsky equation admits special solutions known as algebraically special modes. It would therefore be interesting to determine whether analogous solutions exist for accelerating and rotating black holes. Another important question concerns the relations between perturbations of different spin weights. In the Kerr spacetime, the Teukolsky–Starobinsky identities relate solutions of the Teukolsky equations with opposite spin weights and therefore also provide results regarding isospectrality properties of perturbations with different spin weights. Whether analogous identities exist for the generalized Teukolsky equation in accelerating and rotating spacetimes remains an open question. For a detailed review of algebraically special modes and the Teukolsky–Starobinsky identities, see appendix A of \cite{Berti:2009kk}.

Furthermore, recent work on Kerr and Kerr de Sitter black holes has uncovered hidden spectral symmetries of the Teukolsky equation, providing an alternative way of relating the spectra of perturbations with different spin weights \cite{Casals:2021ugr}. Through these transformations, the complex valued Teukolsky equation can be mapped to a real differential equation, which reduces to the equation introduced by Whiting \cite{Whiting:1989} when the cosmological constant vanishes. The resulting real equation provides a framework for analyzing the mode stability problem, and mode stability has consequently been established for the corresponding Kerr de Sitter problem outside the superradiant regime \cite{Casals:2021ugr}. Given the similarities between the C-metric and de Sitter black holes, it would be interesting to investigate whether a similar construction can be developed for accelerating and rotating black holes. Such a construction could provide a route towards establishing their mode stability, or identifying possible instabilities.

Finally, another interesting path worth exploring is concerned with superradiance instability. The superradiance of charged scalar waves in the charged C-metric has already been analyzed numerically in \cite{Destounis:2022rpk} where it was conjectured that the C-metric is modally stable against charged scalar field perturbations. This is particularly interesting in view of the close similarities between the C-metric and Reissner Nordstr\"{o}m de Sitter black holes. On the other hand, the Reissner Nordstr\"{o}m de Sitter black hole possess regions of parameter where charged scalar perturbations can develop a superradiant instability. Understanding the origin of this difference between the two spacetimes, and whether it can be established analytically, would therefore be an interesting problem for future work.

\paragraph{Acknowledgments:} I am grateful to my supervisor Filipe Moura for discussions and reviewing an early draft of this paper. I would also like to thank David Kofro\v{n} for valuable insights on the separation constant, as well as Gabriel Ramos and Francisco Alves for helpful conversations on the linear stability problem.
\noindent
%%%%%%%%%%%%%%%%%%%%%%%%%%%%%%%%%%%%%%%%%%%%%%%%%%%%%%%%%%%%%%%%
%%%%%%%%%%%%%%%%%%%%%%%%%%%%%%%%%%%%%%%%%%%%%%%%%%%%%%%%%%%%%%%%

\end{document}